\documentclass[lettersize,journal]{IEEEtran}
\ifCLASSOPTIONcompsoc
\usepackage[caption=false, font=normalsize, labelfont=sf, textfont=sf]{subfig}
\else
\usepackage[caption=false, font=footnotesize]{subfig}
\fi
\usepackage{amsmath,amsfonts}
\usepackage{algorithmic}
\usepackage{array}
\usepackage{lipsum}
\usepackage{graphicx}
\usepackage{textcomp}
\usepackage{stfloats}
\usepackage{url}
\usepackage{verbatim}
\usepackage{graphicx}
\usepackage{booktabs}
\usepackage{amssymb}
\usepackage{cite}
\usepackage{bm}
\usepackage{tikz}
\usetikzlibrary{positioning,shapes}
\usepackage{multirow}
\usepackage{siunitx}
\usepackage{makecell}
\usepackage{adjustbox}

\usepackage{booktabs}
\newcommand{\tabitem}{~~\llap{\textbullet}~~}

\def\BibTeX{{\rm B\kern-.05em{\sc i\kern-.025em b}\kern-.08em
		T\kern-.1667em\lower.7ex\hbox{E}\kern-.125emX}}
\usepackage{balance}

\def\ninept{\def\baselinestretch{0.92}\let\normalsize\small\normalsize}

\begin{document}
	\title{Exploration of Learned Lifting-Based Transform \\Structures for Fully Scalable and Accessible\\Wavelet-Like Image Compression}
	\author{Xinyue~Li,~\IEEEmembership{Student Member,~IEEE,} Aous~Naman,~\IEEEmembership{Senior Member,~IEEE,} and David~Taubman,~\IEEEmembership{Fellow,~IEEE}}

\maketitle
	
\begin{abstract} 
	
	This paper provides a comprehensive study on features and performance of different ways to incorporate neural networks into lifting-based wavelet-like transforms, within the context of fully scalable and accessible image compression. Specifically, we explore different arrangements of lifting steps, as well as various network architectures for learned lifting operators. Moreover, we examine the impact of the number of learned lifting steps, the number of channels, the number of layers and the support of kernels in each learned lifting operator. To facilitate the study, we investigate two generic training methodologies that are simultaneously appropriate to a wide variety of lifting structures considered. Experimental results ultimately suggest that retaining fixed lifting steps from the base wavelet transform is highly beneficial. Moreover, we demonstrate that employing more learned lifting steps and more layers in each learned lifting operator do not contribute strongly to the compression performance. However, benefits can be obtained by utilizing more channels in each learned lifting operator. Ultimately, the learned wavelet-like transform proposed in this paper achieves over 25\% bit-rate savings compared to JPEG 2000 with compact spatial support.
	
\end{abstract}
	
\begin{IEEEkeywords}
	image compression, neural networks, wavelet transform, lifting scheme, end-to-end optimization
\end{IEEEkeywords}
	
\section{Introduction}
\label{sec:intro}

In the last decade, learning-based approaches have been successfully developed to improve coding efficiency of image compression applications, with very promising results.
These learning-based approaches are most commonly adopted in an autoencoder framework \cite{balle2016end, minnen2018joint, theis2017lossy, toderici2017full}.
Another class of approaches, which has been studied, involves a learned wavelet-like compression framework \cite{ma2019iwave, dardouri2020optimized, dardouri2021neural, dardouri2021dynamic, Xinyue2021, Xinyue2022, Xinyue2022_journal}; this class of approaches is the focus of this paper.
Wavelet-like compression frameworks particularly lend themselves to the achievement of three important properties:
\begin{itemize}
\item[1.] Resolution scalability
\item[2.] Quality scalability
\item[3.] Random region-of-interest accessibility
\end{itemize}

First of all, a learned wavelet-like compression framework naturally inherits the multi-scale representation from the conventional wavelet transform \cite{akansu2001multiresolution}, which supports reconstruction of an already encoded image at multiple resolutions, a feature known as resolution scalability.
The most common implementation of wavelet-like transforms employs the lifting scheme \cite{sweldens1998lifting,Sweldens1995,SWELDENS1996}, whose idea is to factorize the poly-phase matrix of the wavelet filters into a sequence of elementary convolution steps, known as \emph{lifting steps}.
Conventionally, these lifting steps (known as \emph{predict} and \emph{update} steps) employ fixed filters as the corresponding operators \cite{daubechies1998,daubechies2005factoring}.
The key idea of learned wavelet-like transforms is to replace or augment these fixed lifting operators in the conventional wavelet transform with learned neural networks, while maintaining perfect reconstruction.
More importantly, these wavelet-like transforms are \emph{self-similar}, meaning that the same transform can be uniformly applied to process images at different scales.

In contrast, generic autoencoders employ neural networks as the analysis and synthesis transforms; these networks are usually unstructured, and so do not provide any multi-resolution representation of the image.
As a result, autoencoders do not naturally support resolution scalability.
Even in some cases where the employed neural networks have specific structures, such as U-Net \cite{ronneberger2015u}, the image is still not processed using the same transform at each resolution; in other words, autoencoders do not support the self-similarity feature.

Secondly, to achieve quality scalability, wavelet-like compression frameworks design the transforms to be critically sampled; as a result, an image can be reconstructed at an arbitrary high quality by only incorporating more bits to the bit-stream, a feature known as quality scalability.
By contrast, autoencoders employ neural networks that usually involve dimension reduction for the purpose of compression.
This means that images simply cannot be represented exactly by these transforms (or networks) in autoencoders, even without quantization, which interferes with quality scalability.

In addition, for a compression system to achieve region-of-interest accessibility, it is crucial to impose constraints on its receptive field; certainly, any element that involves non-local operations, such as global normalisation, should be avoided in the system design.	
These constraints naturally fit in traditional wavelet codecs, and so random region-of-interest accessibility is supported.

In this paper, we focus exclusively on the transform itself, investigating and comparing various transform structures for improved learned wavelet-like compression.
For this very reason, we deliberately choose not to incorporate other tools, such as learned pre-/post-processing strategies \cite{aytekin2018block, akyazi2019learning}, hyperpriors \cite{balle2018variational, hu2020coarse, brand2021rate, lin2020spatial} and/or collections of rich context models \cite{lee2018context, li2020efficient, he2021checkerboard, he2022elic}, in this study.
Although these tools can have beneficial impacts on the coding efficiency of learned wavelet-like compression, they can obscure the contribution made solely by the transform itself, and may easily wind up being global and not self-similar, undermining the scalability and accessibility of the compression framework.

Within this context, Ma et al. propose an \emph{iWave} transform \cite{ma2019iwave}; this transform implements the update step first via a simple averaging operation, while the following predict step utilises a Convolutional Neural Network (CNN) as the operator.
The iWave transform improves energy compaction compared to the CDF 9/7-based wavelet transform \cite{CDF_9_7} of the JPEG2000 standard \cite{j2kbook}.
Interestingly, the iWave++ transform developed by some of the same authors \cite{ma2020end}, relies upon a more conventional predict-update lifting structure, extending the utilisation of neural networks to both the predict and update steps.
In \cite{dardouri2020optimized}, Dardouri et al. also propose to employ learned predict and update operators, but with Fully Connected Neural Networks.
This work is further extended in \cite{dardouri2021neural} and \cite{dardouri2021dynamic}; however, improvements over JPEG2000 could be obtained only for the SSIM metric and less well-known PieAPP metric.

In contrast to these existing learned wavelet-like transforms, which replace some or all fixed lifting steps in the wavelet transform with neural networks, our previous works in \cite{Xinyue2022,Xinyue2022_journal} take a different approach.
Specifically, we proposed to augment existing base wavelet transforms that already work well for compression, with two additional learned lifting steps, named \emph{high-to-low} and \emph{low-to-high} steps.
The high-to-low step suppresses aliasing information in the low-pass band using the detail bands at the same level of decomposition, while the following low-to-high step further reduces redundancy in the detail bands using the corresponding low-pass band.

Rather than adopting commonly used CNN or FCNN topologies, the network architecture that we developed for the high-to-low and the low-to-high operators is inspired and guided by a specific theoretical argument related to geometric flow \cite{Xinyue2022_journal}.
This specific network topology involves a collection of purely linear filters modulated by a shallow non-linear sub-network, so that the overall network architecture has limited non-linearities, low computational complexity and relatively small region of support. We used the term \emph{proposal-opacity} to describe this specific network architecture. 

To train the high-to-low and low-to-high steps in an end-to-end optimization framework, we developed a backward annealing approach to model the discontinuous quantization and entropy coding processes during back-propagation \cite{Xinyue2022_journal}.
By learning only one set of network parameters for all levels of decomposition and for all bit-rates of interest, our proposed method in \cite{Xinyue2022_journal} achieves significant bit-rate savings compared to JPEG2000 over a wide range of bit-rates, and also improves visual quality of the reconstructed images at reduced resolutions.
Significantly, the resulting coding framework retains the resolution scalability, quality scalability and region of interest accessibility of JPEG 2000, though the learning-based wavelet-like transform has a larger region of support than the conventional wavelet transform.
Another remarkable difference between our proposed coding framework in \cite{Xinyue2022_journal} and most existing compression schemes is that we encode the image only once to high quality; all reduced-quality representations of the image are derived from this one bit-stream by discarding irrelevant bits.

Despite the success of the aforementioned works, there has been very limited comparative study of the different methods proposed so far \cite{Xinyue2023}, so as to gain a full picture on what is the most effective way to develop learned wavelet-like transforms with the aid of neural networks, for fully scalable and accessible image compression.
Therefore in this paper, we first set out to understand more clearly the value brought by different lifting structures, as well as the merits of various network architectures for learned lifting operators in each lifting structure.

Subsequently, we study the relationship between the depth of lifting structures (i.e. the number of learned lifting steps) and coding performance.
Ultimately, all learned lifting steps in this paper are used in a critically sampled highly scalable compression system, in which only one trained model is applied to all levels of decomposition and all bit-rates of interest.
This is a very different experimental context to that used by most deep learning-based methods for image compression. 

Furthermore, we explore the impact of three particular parameters related to the neural network architecture itself on coding performance -- the number of channels, the number of layers and kernel support in each layer.
Increasing any of these parameters can potentially lead to higher coding efficiency; however, they all affect computational complexity, and the latter two have adverse impact on region of support of the overall compression scheme.
Therefore, it is important to understand which of these parameters contributes the most to compression efficiency, within the constraint of features that we would like to retain from wavelet-like compression.

Ultimately, to draw meaningful conclusions from our study, we explore two robust and effective training strategies in Section~\ref{sec:learning_framework_strategy}, which is applicable to a wide variety of lifting structures that are investigated in this paper.
The rest of this paper is arranged as follows. We first introduce the lifting structures that we choose to investigate in this paper in Section~\ref{sec:investigated_lifting_structure}.
Subsequently, we explain how the proposal-opacity network topology proposed in our previous work \cite{Xinyue2022_journal} is leveraged and extended to the work in this paper in Section~\ref{sec:proposal_opacity_topology}.
Experimental results are shown in Section~\ref{sec:experimental_results}, followed by conclusion and recommendations in Section~\ref{sec:conclusion}.

\section{Investigated lifting structures}
\label{sec:investigated_lifting_structure}
This section introduces the lifting structures investigated in this paper. 
These structures help us understand the value brought by different orders of lifting steps, as well as the impact of employing different numbers of learned lifting steps.

\subsection{Predict-update lifting structure}
\label{subsec:predict-update}
\begin{figure}[htb]
	\centering
	\subfloat[]{%
	\includegraphics[width=\linewidth]{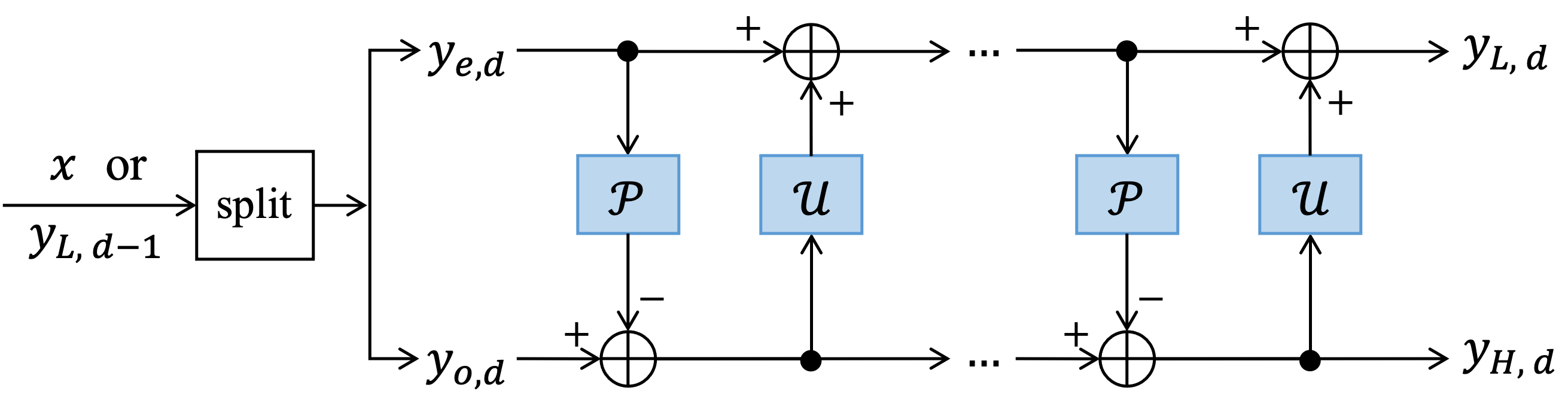}}\\
	\subfloat[]{%
	\includegraphics[width=\linewidth]{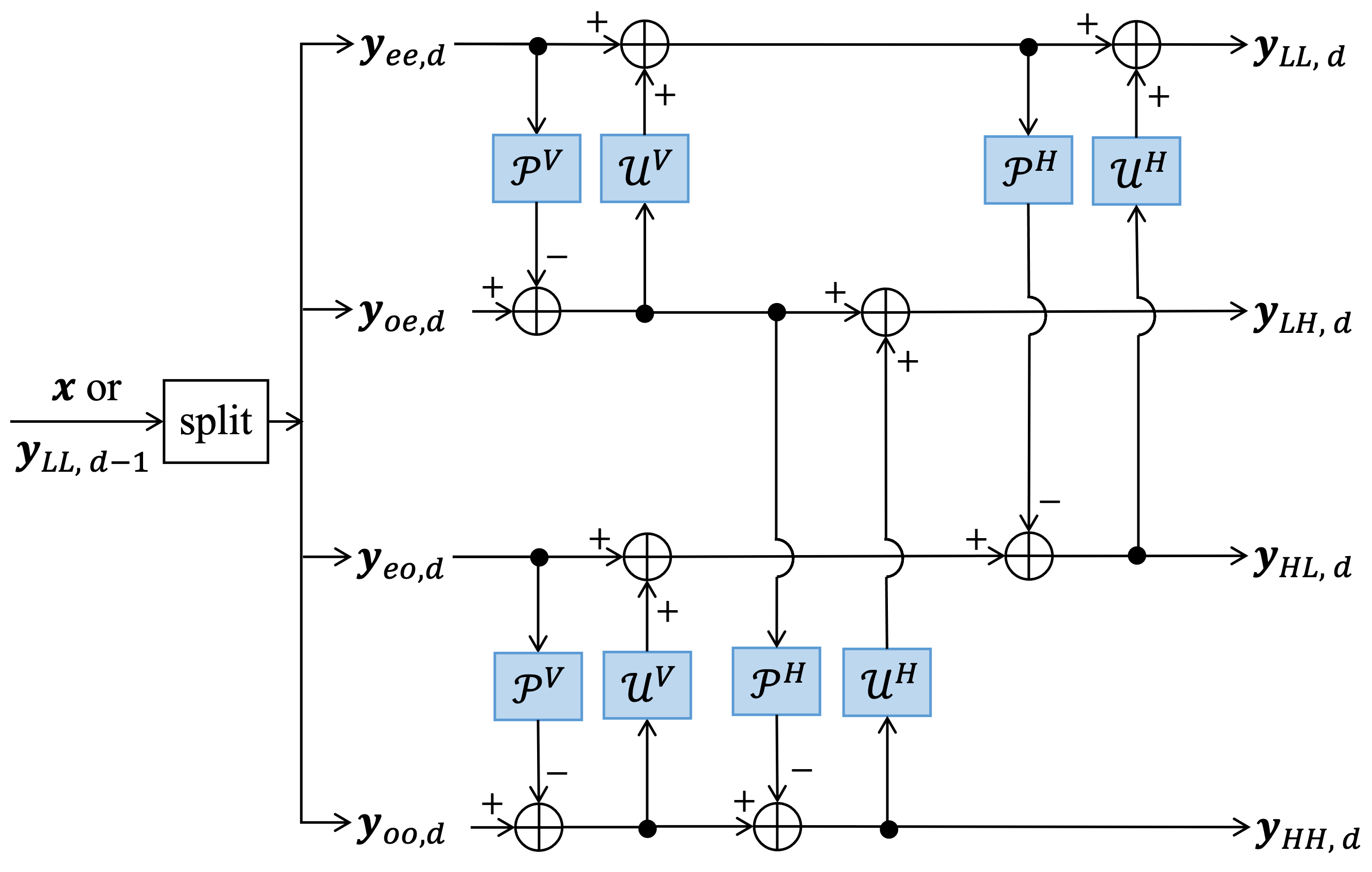}}
	\caption{(a) The general predict-update lifting structure in one dimension, where $\mathcal{P}$ and $\mathcal{U}$ denote the predict and the update operators, respectively. (b) An example of two-dimensional predict-update lifting structure, in which only two steps are shown for horizontal and vertical directions; this is essentially the lifting structure of the LeGall 5/3 wavelet transform. The symbols $\mathcal{P}^{V}$, $\mathcal{U}^{V}$, $\mathcal{P}^{H}$ and $\mathcal{U}^{H}$ denote the vertical-predict, vertical-update, horizontal-predict and horizontal-update operators, respectively.}
	\label{fig:predict_update}
\end{figure}

For the conventional wavelet transform, the most commonly adopted lifting scheme has a \emph{predict-update} structure, as illustrated in Fig.~\ref{fig:predict_update}(a).
This lifting structure first predicts the odd samples $y_{o, d}$ using the even samples $y_{e, d}$ through a predict operator $\mathcal{P}$.
Subsequently, the even samples $y_{e, d}$ are updated, using the update operator $\mathcal{U}$, which operates on the odd samples produced by the predict step.
These two steps alternate one or more times to produce the low-pass and the high-pass bands, $y_{L, d}$ and $y_{H, d}$. 

This predict-update structure can be easily extended to build two-dimensional wavelet transforms, cascading a number of predict and update steps in both horizontal and vertical directions.
A simple example of this is the LeGall 5/3 wavelet transform; it employs two lifting steps (one predict and one update) in both horizontal and vertical directions, as illustrated in Fig.~\ref{fig:predict_update}(b).
The lifting operators $\mathcal{P}^{V}$, $\mathcal{U}^{V}$, $\mathcal{P}^{H}$ and $\mathcal{U}^{H}$ in the LeGall 5/3 wavelet transform are simply one-dimensional filters with transfer functions as
\begin{align}
\mathcal{P}(z) &= -\frac{1}{2}(1 + z) \label{eq:P}\\
\mathcal{U}(z) &= \frac{1}{4}(1 + z^{-1}) \label{eq:U}
\end{align}
and so
\begin{align}
\mathcal{P}^{V}(z_1, z_2) = \mathcal{P}(z_1)&,~~ \mathcal{P}^{H}(z_1, z_2) = \mathcal{P}(z_2) \\
\mathcal{U}^{V}(z_1, z_2) = \mathcal{U}(z_1)&,~~ \mathcal{U}^{H}(z_1, z_2) = \mathcal{U}(z_2)
\end{align}
There are other wavelet transforms which employ more lifting steps within the predict-update structure; for instance, the CDF 9/7 wavelet transform employs four lifting steps (two predict and two update) in each direction.
Explorations on employing more learned lifting steps to build wavelet-like transforms will be provided shortly.
At this stage, we restrict our attention to the case shown in Fig.~\ref{fig:predict_update}(b).

As we have mentioned earlier, the key idea of learned wavelet-like transforms is to replace the fixed lifting operators in the conventional wavelet transform with neural networks.
Therefore in this paper, we replace each of the operators $\mathcal{P}^{V}$, $\mathcal{U}^{V}$, $\mathcal{P}^{H}$ and $\mathcal{U}^{H}$ individually with a learned two-dimensional neural network.

\subsection{Update-predict lifting structure}
\label{subsec:update-predict}
\begin{figure}[htb]
	\centering
	\includegraphics[width=\linewidth]{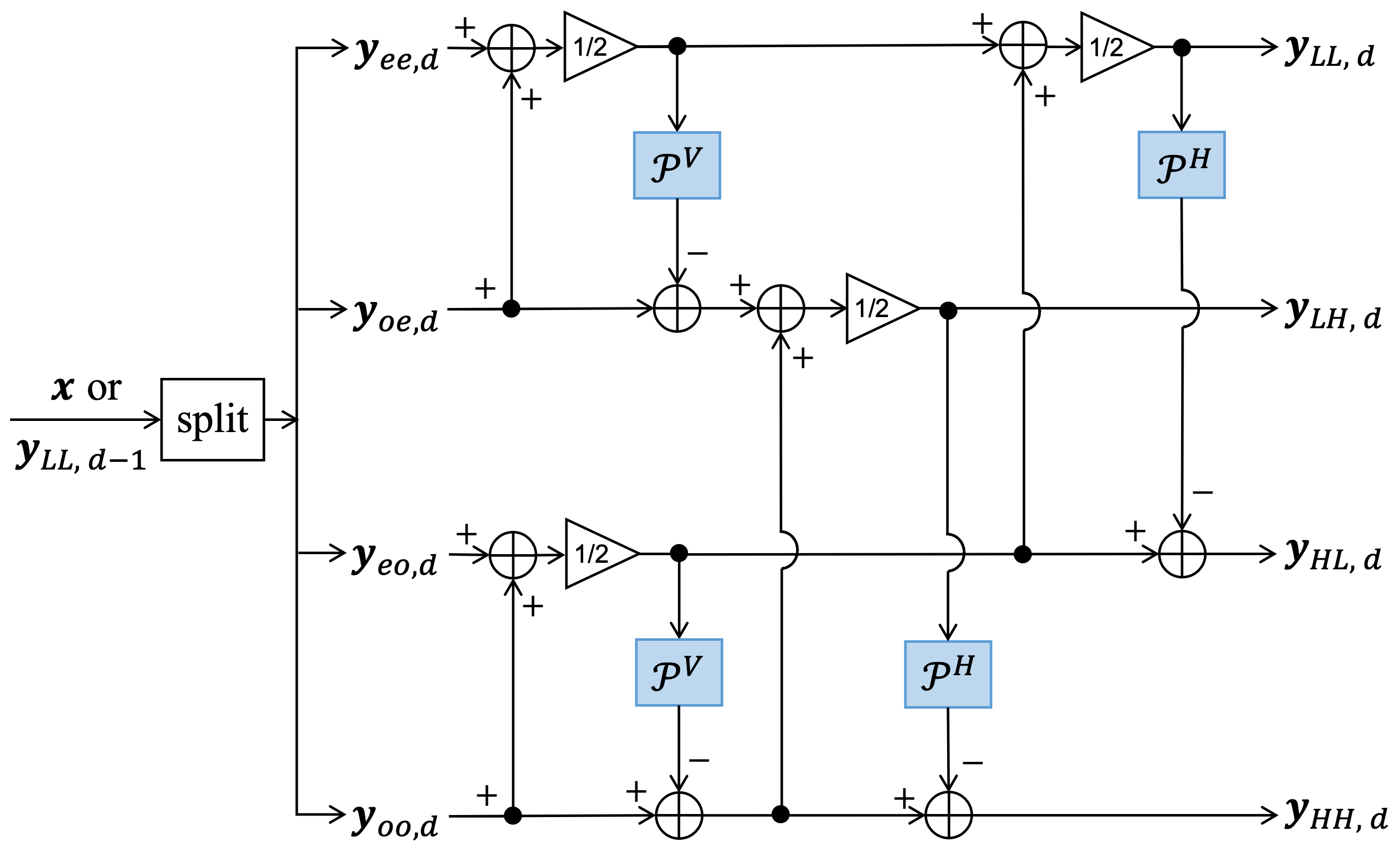}
	\caption{The update-predict lifting structure explored in this paper; this structure was first proposed in \cite{claypoole2003nonlinear}, and was also employed to build the iWave transform in \cite{ma2019iwave}.}
	\label{fig:update_predict_LS}
\end{figure}

In the literature, update-predict lifting structures have also been considered, particularly in the context of developing adaptive lifting operators \cite{claypoole2003nonlinear}.
In fact, \cite{ma2019iwave} demonstrates that the update-predict structure is more favourable than the predict-update structure for learned wavelet-like compression; such kind of structure is illustrated in Fig.~\ref{fig:update_predict_LS}, where a simple averaging operation is employed for the update steps, as proposed in \cite{ma2019iwave}.
However, the work presented in \cite{ma2019iwave} does not involve a full end-to-end optimization that includes quantization and entropy coding during training.

\subsection{Hybrid lifting structure}
\label{subsec:hybrid_structure}
In addition to the predict-update and update-predict structures, other possibilities exist, in which only some lifting steps involve neural networks.
An excellent example is the \emph{hybrid} structure that we introduce here, which starts with a fixed base wavelet transform, and augments it with two additional learned lifting steps, named the high-to-low and low-to-high steps.

\begin{figure}[htb]
	\centering
	\includegraphics[width=\linewidth]{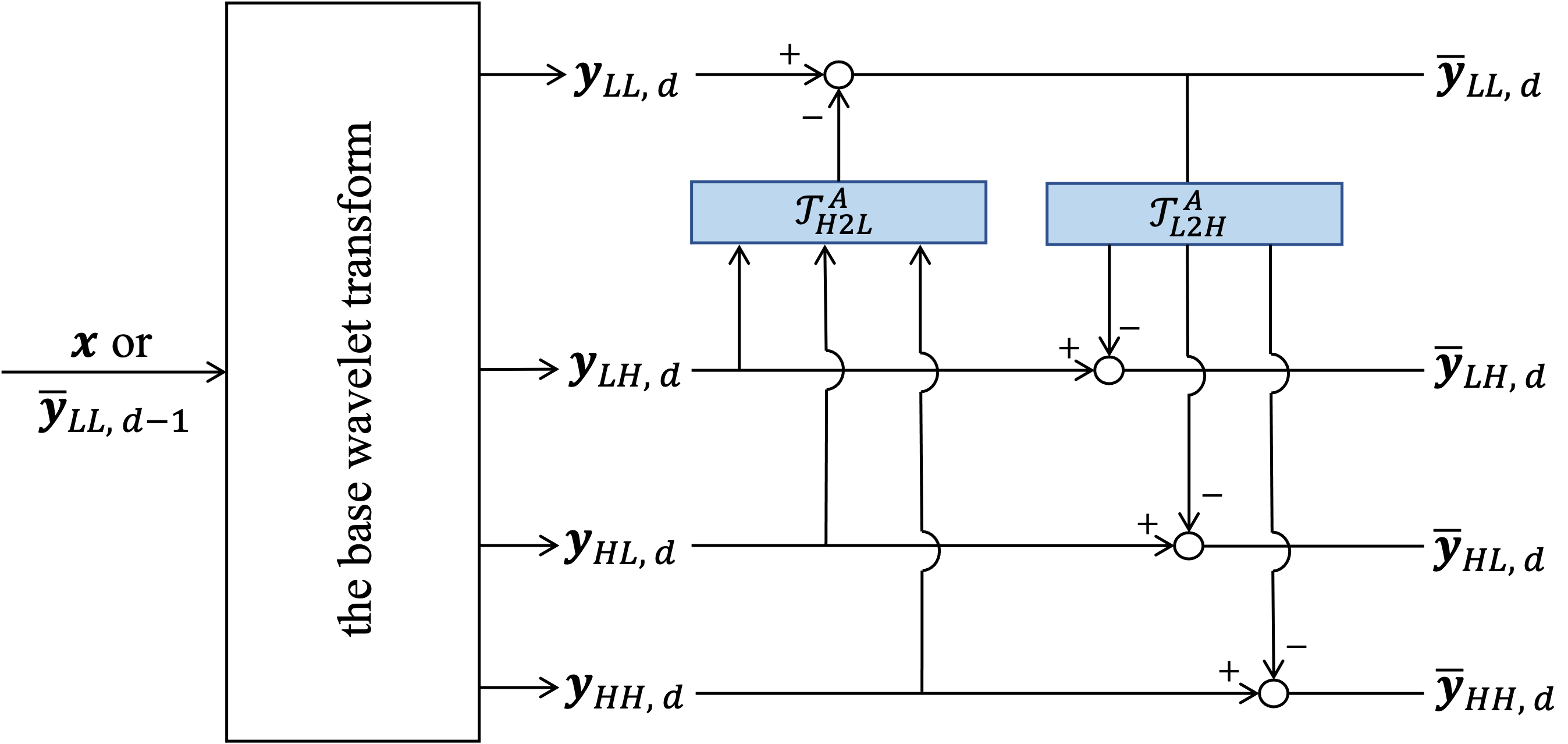}
	\caption{The hybrid lifting structure introduced in this paper; $T^{A}_{H2L}$
and $T^{A}_{L2H}$ denote the high-to-low and the low-to-high operators using neural networks.}
	\label{fig:hybrid_LS}
\end{figure}

The high-to-low step aims to predict redundant information in the wavelet low-pass subband, utilizing the detail bands at the same resolution, as seen in Fig.~\ref{fig:hybrid_LS}; we do this using the learned operator $T^{A}_{H2L}$.
This high-to-low step can effectively clean the accumulation of aliasing through the DWT hierarchy, which improves visual quality of the reconstructed images at different scales and coding efficiency, as demonstrated in \cite{Xinyue2022_journal}.

The following low-to-high step further compacts the high-frequency coefficients of the wavelet transform, using the corresponding \emph{cleaned} low-pass band; we do this by employing the learned operator $T^{A}_{L2H}$.
Besides the benefit of improving coding efficiency, this auxiliary low-to-high step also helps reduce the dependence of the high-to-low step on high quality quantized detail subbands during reconstruction.
Even if all of the encoded detail subband samples wind up being quantized to zero, the original detail samples can still be partially restored by the low-to-high step during wavelet synthesis, after which the high-to-low step is able to at least partially restore the aliasing correction term at the decoder.
As a result, the combination of the high-to-low and the low-to-high steps effectively introduces super-resolution into the wavelet synthesis process.

It is worthwhile to mention that unlike the predict-update and update-predict structures, which cascade horizontal and vertical learned processes in a particular order, this hybrid structure has the benefit of treating horizontal and vertical directions equally.

\subsection{Lifting structures with more learned lifting steps}
\label{subsec:LS_more_steps}
As we have seen in Section~\ref{subsec:hybrid_structure}, the hybrid lifting structure essentially comes with more lifting steps, but fewer of them involve learning -- fixed lifting steps in the existing base wavelet transform plus only two learned lifting steps.
Of course, the question remains if neural networks can be beneficially employed in the lifting steps that correspond to the base wavelet transform as well.

\begin{figure}[htb]
	\centering
	\includegraphics[width=\linewidth]{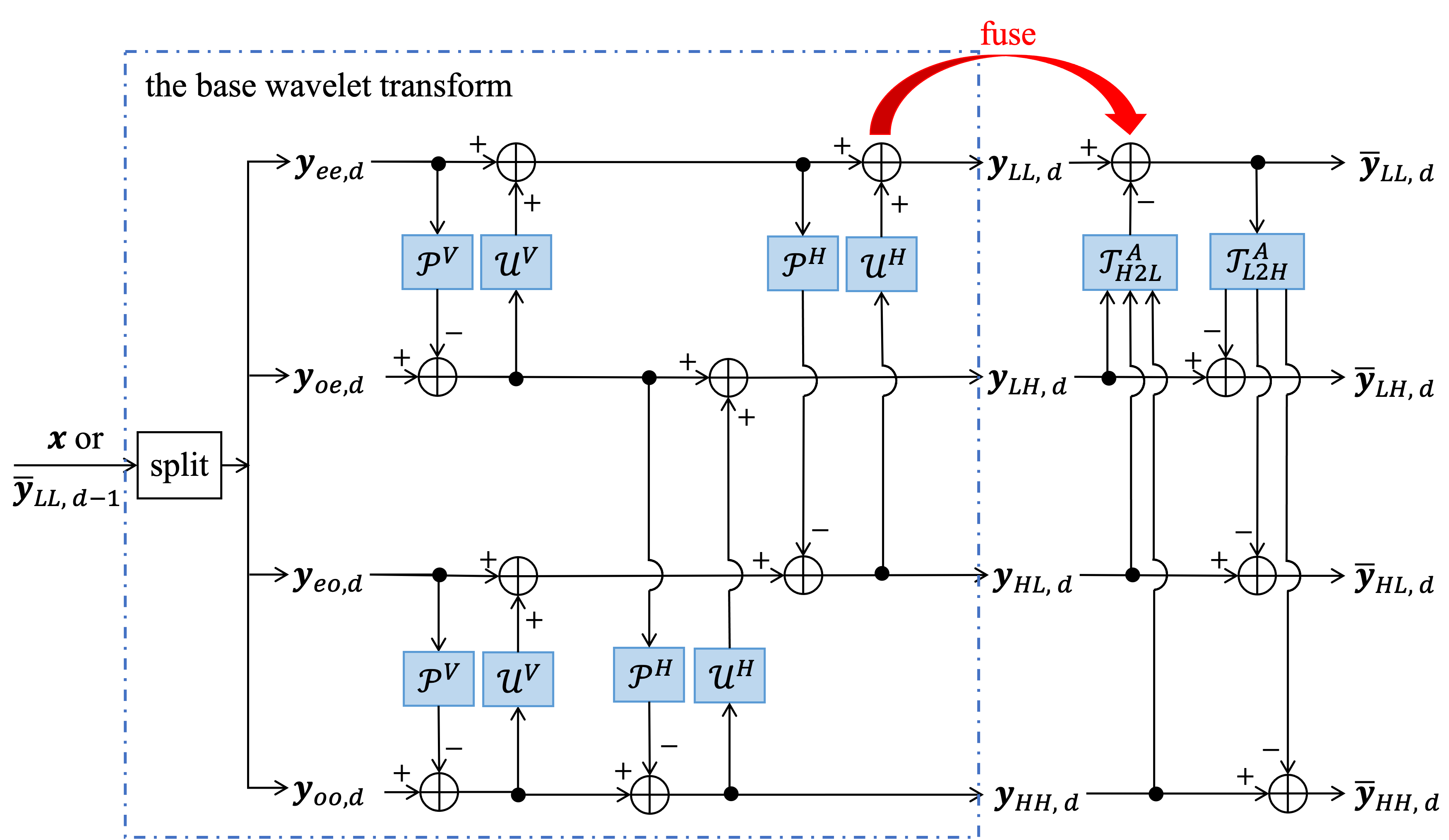}
	\caption{The extension of neural networks to the base wavelet transform shown in Fig.~\ref{fig:hybrid_LS}. Note that the last update step $\mathcal{U}^H$ can be fused into the high-to-low step $\mathcal{T}^H_{H2L}$.}
	\label{fig:Custom_5S}
\end{figure}

To address this, we factorize the LeGall 5/3 base wavelet transform into a sequence of lifting steps as seen in Fig.~\ref{fig:Custom_5S}, each of which can then be replaced individually with a learned equivalent.
In addition, when we put these learned lifting networks together with the two additional steps $\mathcal{T}^A_{H2L}$ and $\mathcal{T}^A_{L2H}$, the resulting lifting structure in Fig.~\ref{fig:Custom_5S} can also be capable of discovering the solution presented in Section~\ref{subsec:hybrid_structure}.
More importantly, since both $\mathcal{U}^H$ and $\mathcal{T}^A_{H2L}$ are update operators, they can be fused together as highlighted in Fig.~\ref{fig:Custom_5S}, so that the total number of learned lifting steps can be reduced by one.

It is worthwhile to point out that the replacement of each individual fixed lifting step in the base wavelet transform with a neural network comes with a cost.
This is because each learned lifting operator/network generally has a substantially larger region of support as well as higher computational complexity than the corresponding fixed lifting operator.
More importantly, since each learned lifting operator/network usually involves non-linearities, quantization errors can expand in an uncontrollable way through these non-linearities during synthesis.
This is particularly important for the work in this paper, because we aim to employ only one single trained model of learned lifting operators for all levels of the wavelet decomposition and for all bit-rates of interest over a wide range, leading to a highly scalable compression system that preserves quality scalability, resolution scalability and region-of-interest accessibility.

Furthermore, a less obvious consequence of increasing the number of learned lifting steps is that the training becomes very difficult.
In fact, as we shall see in Section~\ref{subsubsec:significance_pre_training}, these learned steps can easily wind up exhibiting worse coding efficiency than the base wavelet transform.

\section{Investigated Network Topologies}
\label{sec:proposal_opacity_topology}
In this section, we introduce a particular \emph{proposal-opacity} network topology, which can be applied uniformly to construct the learned lifting operators in Section~\ref{sec:investigated_lifting_structure}.
Later, we compare it with two other well-known network architectures for learned wavelet-like transforms -- \emph{iWave} and \emph{iWave++} topolgies as proposed in \cite{ma2019iwave} and \cite{ma2020end}.

%

\subsection{Significance of the proposal-opacity topology}
\label{subsec:significance_proposal_opacity_topology}
The proposal-opacity network topology was originally advanced in our work \cite{Xinyue2022_journal}, in the context of the hybrid architecture.
As explained earlier, the high-to-low and low-to-high steps serve to remove residual redundancy between successive levels of the wavelet transform; a major form of this residual redundancy comes as aliasing information in the wavelet transform. 

\begin{figure}[htb]
	\centering
	\includegraphics[width=1.0\linewidth]{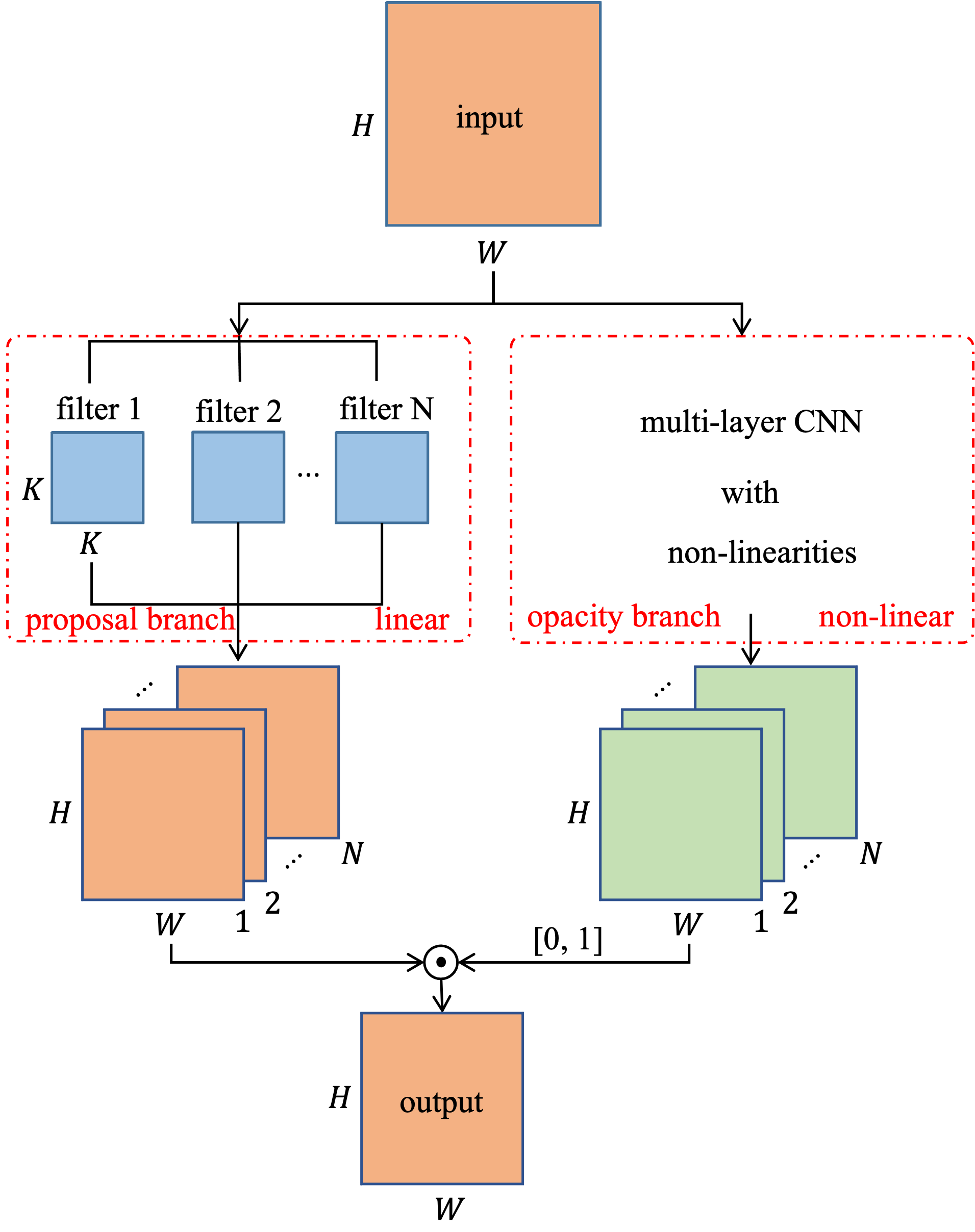}
	\caption{The proposal-opacity neural network topology proposed in our previous work \cite{Xinyue2022_journal}. The symbol $K \text{~x~} K$ denotes the filter support while $N$ represents the number of filters (or equivalently the number of channels in the proposal/opacity branch).}
	\label{fig:proposal_opacity_topology}
\end{figure}

In fact, our previous work in \cite{Xinyue2022_journal} has shown that if we know local orientations (i.e. geometric flow) a priori, then the solution to eliminating aliasing information from both the low- and the high-pass subbands of the wavelet transform can be simply a linear filter.
However, since geometric flow is a local property that is hard to accurately determine within an image, in reality, the untangling of aliasing from the wavelet subbands requires either an adaptive filtering solution or a bank of filters with an adaptive strategy for combining their responses.
The latter one is exactly the proposal-opacity structure that we introduce here -- the proposals essentially form a bank of linear filters (or can be understood as candidate lifting steps), while the opacities provide a data-dependent blending of these linear proposals, as depicted in Fig.~\ref{fig:proposal_opacity_topology}.

Additionally, it turns out that the proposal-opacity topology employing linear proposals is superior to other topologies that involve more complex non-linear networks for the proposals.
As we shall see in Section~\ref{subsec:proposed_lifting_networks}, this is also the network architecture that we decide to leverage in this paper, when studying the extension of neural networks to all the lifting steps $\mathcal{P}^{V}$, $\mathcal{U}^{V}$, $\mathcal{P}^{H}$ and $\mathcal{U}^{H}$ within the base wavelet transform.

More importantly, this proposal-opacity network topology offers the benefit that it can easily replicate the fixed lifting filters in the base wavelet transform. For instance, if the $N$ filters in the proposal branch are all fixed as the one in \eqref{eq:P}, then the proposal-opacity network always produces the fixed lifting filter in \eqref{eq:P} regardless of the outcomes of the opacity branch.
This additional benefit is important for developing learned networks for $\mathcal{P}^{V}$, $\mathcal{U}^{V}$, $\mathcal{P}^{H}$ and $\mathcal{U}^{H}$, as we expect the learned lifting steps to be still capable of discovering the base wavelet transform as one possible solution during training, which already works well in terms of coding efficiency.

\subsection{Particular properties of the opacity branch}
\label{subsec:property_opacity_branch}

Since the opacity branch in Fig.~\ref{fig:proposal_opacity_topology} can be understood as analysing local scene geometry to produce opacities (or likelihoods) that are used to blend linearly generated proposals, it is reasonable to constrain the outcomes of the opacity branch to be non-negative and summed up to $1$.
Moreover, an important pre-requisite for the opacity branch is that it is capable of producing opacity maps that are invariant to absolute image intensity and contrast.

\begin{figure}[htb]
	\centering
	\subfloat[]{%
	\includegraphics[width=0.4\linewidth]{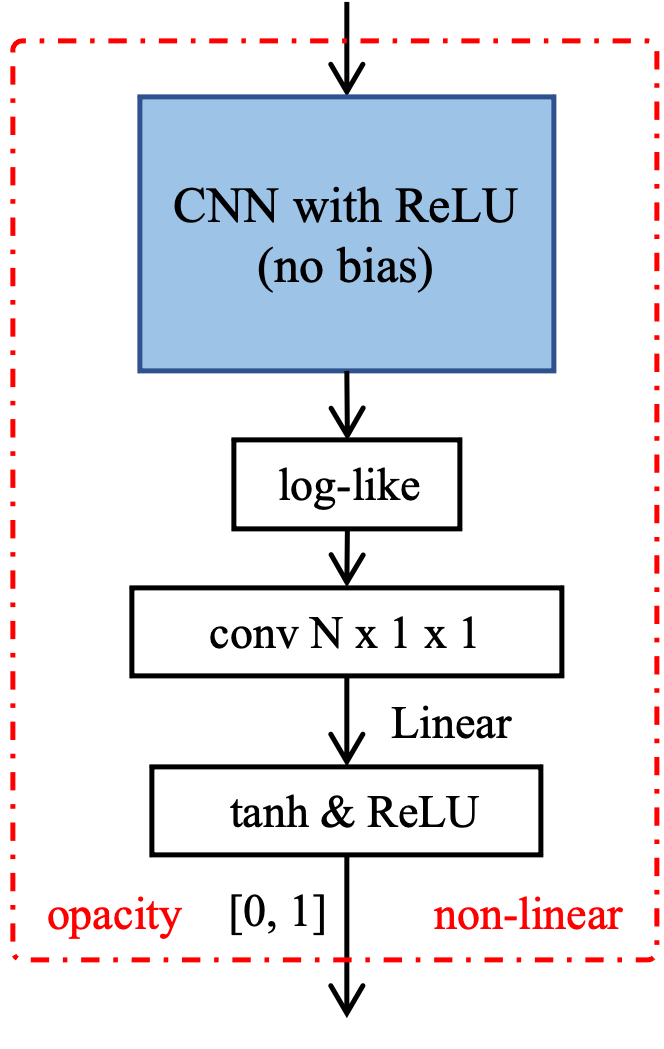}}\hfill
	\subfloat[]{%
	\includegraphics[width=0.4\linewidth]{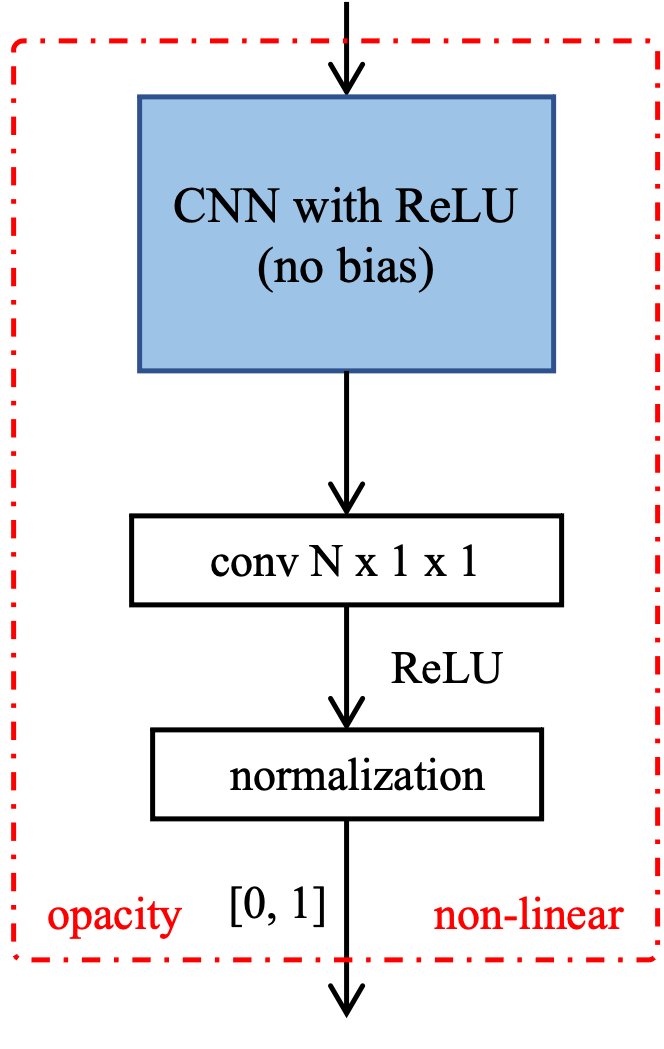}}
	\caption{(a) depicts the log-like activation function after a succession of convolutional layers (no bias) with $ReLU$; this is the opacity architecture that we adopt in our previous work \cite{Xinyue2022_journal}. (b) illustrates the normalization block that we employ in this paper to force the intensity-contrast independence of the opacity branch shown in Fig.~\ref{fig:proposal_opacity_topology}. The notation \emph{conv} $N$ x $K$ x $K$ represents the convolutional layer with $N$ channels (or filters) and kernel support $K \text{~x~} K$.}
	\label{fig:activation_blocks}
\end{figure}

In practice, we achieve these by using a normalization block as shown in Fig.~\ref{fig:activation_blocks}(b).
The normalization function employed here is defined as
\begin{align}
y = \frac{x_i + \text{offset}}{\sum_{i=1}^{N}(x_i + \text{offset})}, \text{~~offset} = 0.01
\end{align}
which is also capable of removing arbitrary intensity scaling factors from the input data.
This is slightly different to the opacity network that we proposed in \cite{Xinyue2022_journal}, which does not come with the property of having all opacities summed up to $1$, as shown in Fig.~\ref{fig:activation_blocks}(a).

The design of Fig.~\ref{fig:activation_blocks}(a) is acceptable for the approach explored in \cite{Xinyue2022_journal}, because the two additional learned operators $\mathcal{T}^A_{H2L}$ and $\mathcal{T}^A_{L2H}$ merely modify the behaviour of the base wavelet transform.
Therefore, it is not fundamentally problematic if all opacities turn out to be zero in \cite{Xinyue2022_journal}.
However, utilizing such an approach may not be appropriate when replacing all lifting operators within the base wavelet transform with neural networks in this paper.
This is because at least one of the proposals needs to be in play in this scenario, which means the opacity branch should not deliver likelihoods that are all close to zero.

\subsection{Proposed lifting networks}
\label{subsec:proposed_lifting_networks}

\begin{figure}[htb]
	\centering
	\includegraphics[width=1.0\linewidth]{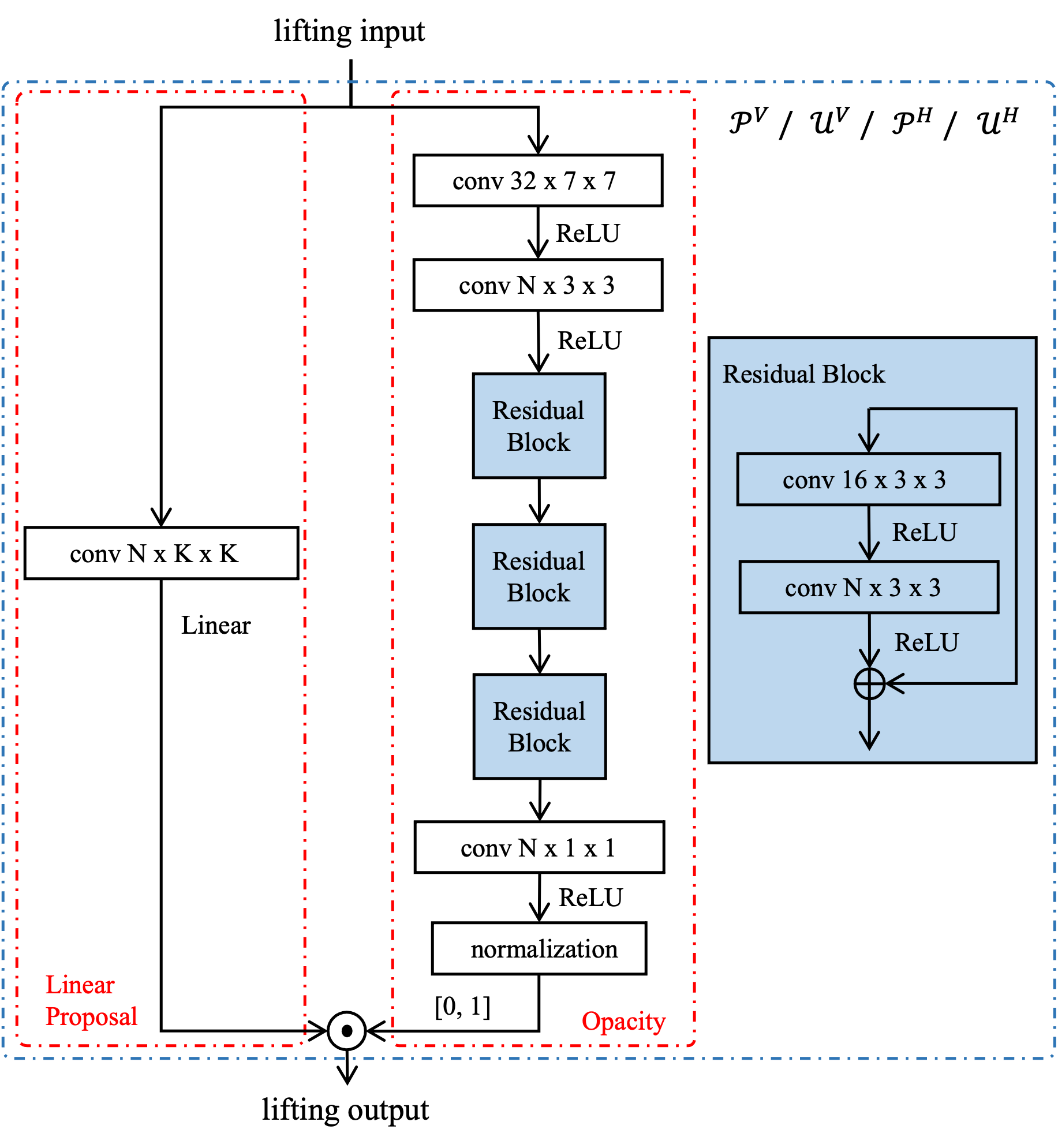}
	\caption{The common proposal-opacity network architecture that is utilized for all the lifting steps $\mathcal{P}^{V}$, $\mathcal{U}^{V}$, $\mathcal{P}^{H}$ and $\mathcal{U}^{H}$ within the base wavelet transform in Fig.~\ref{fig:Custom_5S}. The notation \emph{conv} $N$ x $K$ x $K$ represents the convolutional layer with $N$ channels (or filters) and kernel support $K \text{~x~} K$. The linear proposal branch is chosen to have the same region of support as the opacity branch.}
	\label{fig:proposed_lifting_networks}
\end{figure}

\begin{figure}[htb!]
	\centering
	\subfloat[]{%
	\includegraphics[width=0.76\linewidth]{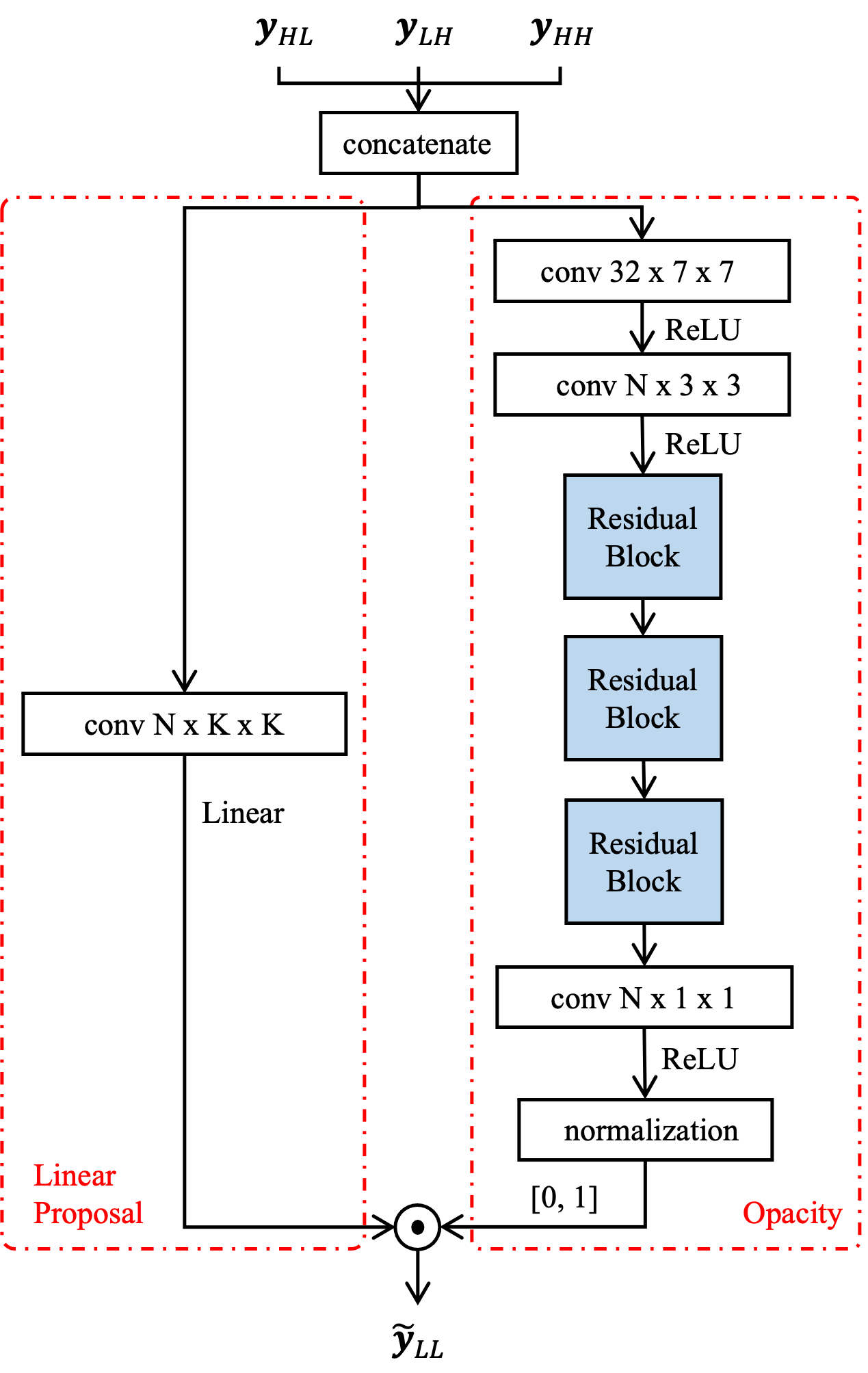}}\\
	\subfloat[]{%
	\includegraphics[width=0.96\linewidth]{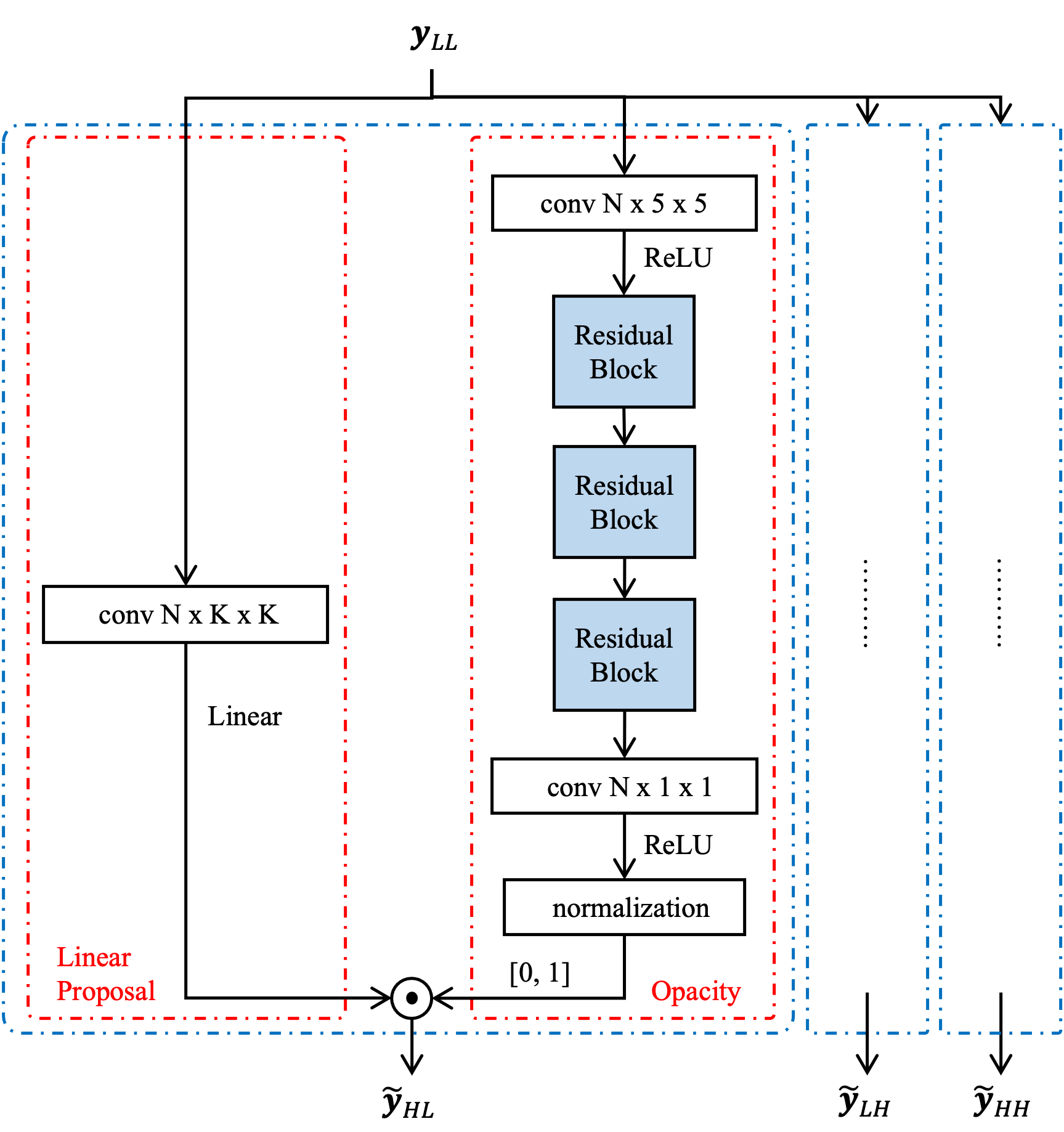}}
	\caption{The high-to-low and the low-to-high networks $\mathcal{T}^A_{H2L}$ and $\mathcal{T}^A_{L2H}$ employed in this paper; the residual blocks have the same structure as in Fig.~\ref{fig:proposed_lifting_networks}. The notation \emph{conv} $N$ x $K$ x $K$ represents the convolutional layer with $N$ channels (or filters) and kernel support $K \text{~x~} K$. The linear proposal branch is chosen to have the same region of support as the corresponding opacity branch. The symbols $\widetilde{\mathbf{y}}_{LL}$, $\widetilde{\mathbf{y}}_{HL}$, $\widetilde{\mathbf{y}}_{LH}$ and $\widetilde{\mathbf{y}}_{HH}$ denote the aliasing information within the low- and high-pass bands.}
	\label{fig:proposed_H2L_L2H_networks}
\end{figure}

In this paper, we apply the proposal-opacity network topology with the general opacity architecture in Fig.~\ref{fig:activation_blocks}(b) to all learned lifting operators $\mathcal{P}^{V}$, $\mathcal{U}^{V}$, $\mathcal{P}^{H}$, $\mathcal{U}^{H}$, $\mathcal{T}^A_{H2L}$ and $\mathcal{T}^A_{L2H}$, as depicted in Fig.~\ref{fig:proposed_lifting_networks} and Fig.~\ref{fig:proposed_H2L_L2H_networks}.
The architecture of the opacity branch is heavily inspired by our previous development in \cite{Xinyue2022_journal}, employing residual blocks that have been
demonstrated to be useful in feature detection for the non-linear opacity branch \cite{li2018learning}.
The linear proposal branch is chosen to have the same region of support as the opacity branch.

Importantly, we employ exactly the same weights of learned lifting operators for all levels of the wavelet decomposition, as well as all bit-rates of interest.
This reinforces the learned wavelet-like transforms to have self-similar interpretation of images, and also supports resolution scalability, quality scalability and random region-of-interest accessibility as mentioned in Section~\ref{sec:intro}.

\section{End-to-end optimisation framework and pre-training strategies}
\label{sec:learning_framework_strategy}
In this section, we introduce the end-to-end optimisation framework to train each learned lifting structure for multiple levels of decomposition; since rate-distortion optimisation objectives are discontinuous, extra care needs to be taken when formulating these objectives for back-propagation.
In addition, just performing the end-to-end optimisation alone does not always lead to good results, as it can easily get trapped into undesirable local optima.
Therefore, we also introduce pre-training strategies in this section, which can help to encourage the convergence to good solutions.

\subsection{End-to-end optimisation with backward annealing}
\label{subsec:end2end_learning_strategy}
In the end-to-end optimisation framework, we consider selectively including an aliasing suppression term as part of the rate-distortion training objective, which is given by
\begin{equation}
\resizebox{\linewidth}{!}{%
$J(\bm{\phi}) = \underbrace{\|\mathbf{x} - \mathbf{\hat{x}(\bm{\phi})}\|^2}_{D} + \lambda_1 \underbrace{\sum_{\beta} \sum_{i \in B_{\beta}} l_{i, \beta}}_{L} + \lambda_2 \underbrace{\sum_d \left\| \widetilde{\mathbf{y}}_{LL, d} (\bm{\phi}) - \widetilde{\mathbf{y}}^t_{LL, d} \right\|_{2}^2}_{\text{aliasing constraint term}}$}
\label{eq:R_D}
\end{equation}
where
\begin{align}
l_{i, \beta} = \log_2 \frac{1}{P_{V_{\beta}}(q_{i, \beta}; \bm{\phi})}= \log_2 \frac{1}{\text{Prob} (V_{\beta}=q_{i, \beta}; \bm{\phi})}
\label{eq:l_b}
\end{align}
In \eqref{eq:R_D}, the total distortion term $D$ represents the sum of squared errors between the input image $\mathbf{x}$ and its reconstructed counterpart $\mathbf{\hat{x}}$; $\bm{\phi}$ represents the vector of all network weights.
The total coded length term $L$ is the sum of all coded lengths $l_{i, \beta}$, resulting from the coding of quantization indices $q_{i, \beta}$ for all subbands $B_{\beta}$.
We use random variable $V_{\beta}$ to model the statistics of the quantization indices $q_{i, \beta}$; then, the coded length $l_{i, \beta}$ can be formulated by \eqref{eq:l_b}.
The aliasing constraint term in \eqref{eq:R_D} measures the sum of squared errors between $\widetilde{\mathbf{y}}_{LL, d}$ and $\widetilde{\mathbf{y}}^t_{LL, d}$ at level $d$ of the wavelet decomposition; $\widetilde{\mathbf{y}}_{LL, d}$ is the aliasing prediction from the high-to-low step $\mathcal{T}^A_{H2L}$, while $\widetilde{\mathbf{y}}^t_{LL, d}$ denotes the target aliasing model constructed in \cite{Xinyue2022_journal}.

The Lagrange multiplier $\lambda_1$ controls the trade-off between distortion $D$ and coded length $L$, while the other Lagrange multiplier $\lambda_2$ controls the level of emphasis on visual quality of the reconstructed images at different scales.
In our previous work \cite{Xinyue2022_journal}, we have demonstrated that $\lambda_2$ can have a beneficial impact on perceptual quality of intermediate low-resolution images across different scales, without significant loss in coding efficiency.
To simplify the experimental conditions of this paper, which focus exclusively on coding efficiency, we stick to the case where $\lambda_2 = 0$.

As recognised by existing works in the literature, end-to-end training targeting the objective in \eqref{eq:R_D} requires a good strategy to model the quantization and the entropy coding processes, which are both discontinuous.
For this purpose, we choose to adopt the end-to-end optimization framework developed in \cite{Xinyue2022_journal} for the work in this paper.
This particular end-to-end optimization framework employs a backward annealing approach, which has certain advantages over additive noise approaches \cite{balle2016end, agustsson2020universally}, the straight-through estimator \cite{bengio2013estimating, yin2019understanding} and soft-to-hard annealing approaches \cite{agustsson2017soft, yang2020improving}, as we have elaborated in \cite{Xinyue2022_journal}.

\subsection{Investigated pre-training strategies}
\label{subsec:pre_training}
The pre-training stage is particularly crucial for the work in this paper for two reasons.

The first reason is because the lifting structure in Fig.~\ref{fig:Custom_5S} becomes \emph{unstructured} once we replace all the fixed lifting steps that correspond to the base wavelet transform with neural networks.
If we employ random initialization for all learned lifting steps in this paper, we lose connection to the base wavelet transform that otherwise exits in \cite{Xinyue2022_journal}, which serves as a stable starting point for training.
As a result, instead of improving upon the conventional base wavelet transform, we can easily wind up with coding performance that is below the conventional base wavelet transform, regardless of the employed network architectures, as we shall see in Section~\ref{subsubsec:significance_pre_training}.

The second issue, which highlights the importance of a good pre-training strategy for the work in this paper, arises from the proposal-opacity topology shown in Fig.~\ref{fig:proposal_opacity_topology}. 
The multiplicative operator makes this structure particularly prone to the vanishing gradient problem \cite{hochreiter1998vanishing}; for instance, if any of the opacities becomes zero during the optimisation procedure, then the corresponding proposal can hardly learn anything useful, and vice versa.
This essentially means that the proposal and the opacity branches are so interdependent that each one ultimately establishes the gradient experienced by the other one during back-propagation.

\subsubsection{Oracle-opacity pre-training schedule}
\label{subsubsec:oracle_training_schedule}
To address these obstacles, we first propose an oracle-opacity pre-training schedule, which consists of three stages as summarised in Table~\ref{table:summary_oracle_opacity_training}. 
Each stage within this particular pre-training schedule is trained using the end-to-end optimization framework with backward annealing as described in Section~\ref{subsec:end2end_learning_strategy}, directly targeting the rate-distortion objective.
\begin{table*}[htb!]
	\caption{The Proposed Oracle-opacity Training Schedule}
	\resizebox{\linewidth}{!}{%
	\centering
	\captionsetup[subfloat]{position=top}
	\begin{tabular}{lllll}
	\toprule
	& \multicolumn{1}{c}{Proposals} & \multicolumn{1}{c}{Opacities} & & \\
	\midrule	
	\multirow{2}{*}{Stage 1: freeze opacities to train proposals} & 
	\makecell{\tabitem one proposal: frozen as the corresponding separable lifting filter \\ (as in either \eqref{eq:P} or \eqref{eq:U})} & 
	\multirow{2}{*}{\makecell{oracle opacities: frozen \\ (obtained off-line)}} & & \\
	& \tabitem the rest $N-1$ proposals: stay trainable & & & \vspace{10pt}\\
	Stage 2: freeze proposals to train opacities & 
	\multicolumn{1}{c}{frozen as trained in Stage 1} &
	trainable from scratch & & \vspace{10pt}\\
	Stage 3: free all proposals and opacities &
	\multicolumn{1}{c}{trainable, starting from Stage 2} & \multicolumn{1}{c}{trainable, starting from Stage 2} & & \\
	\bottomrule
	\end{tabular}}
	\label{table:summary_oracle_opacity_training}
\end{table*}

To be more specific, \emph{Stage 1} derives an initial set of opacities directly from the full-resolution input images, using \emph{heuristics} that extract oriented features.
This initial set of opacities are then fixed, while letting the proposals to learn.
This method is completely impossible in practice, because the decoder does not have access to the full-resolution input images, to recover the same opacities from the inference machine.
For this reason, we refer to these opacities as \emph{oracle opacities}.

Moreover, in \emph{Stage 1} we freeze one of the filters in the proposal branch to have the same coefficients as the corresponding fixed lifting operator (either \eqref{eq:P} or \eqref{eq:U}) in the base wavelet transform, whereas the rest $N - 1$ proposals stay trainable in this stage.
In this way, the lifting network is at least capable of discovering the corresponding fixed lifting operator as one possible solution, which already works well in terms of coding efficiency. 
Subsequently, in \emph{Stage 2} the proposals are frozen, while the opacity branch is allowed to learn from scratch.
In \emph{Stage 3}, we free both the proposal and opacity branches for training, starting from the weights found in \emph{Stage 2}.
\begin{figure}[htb]
	\centering
	\includegraphics[width=0.7\linewidth]{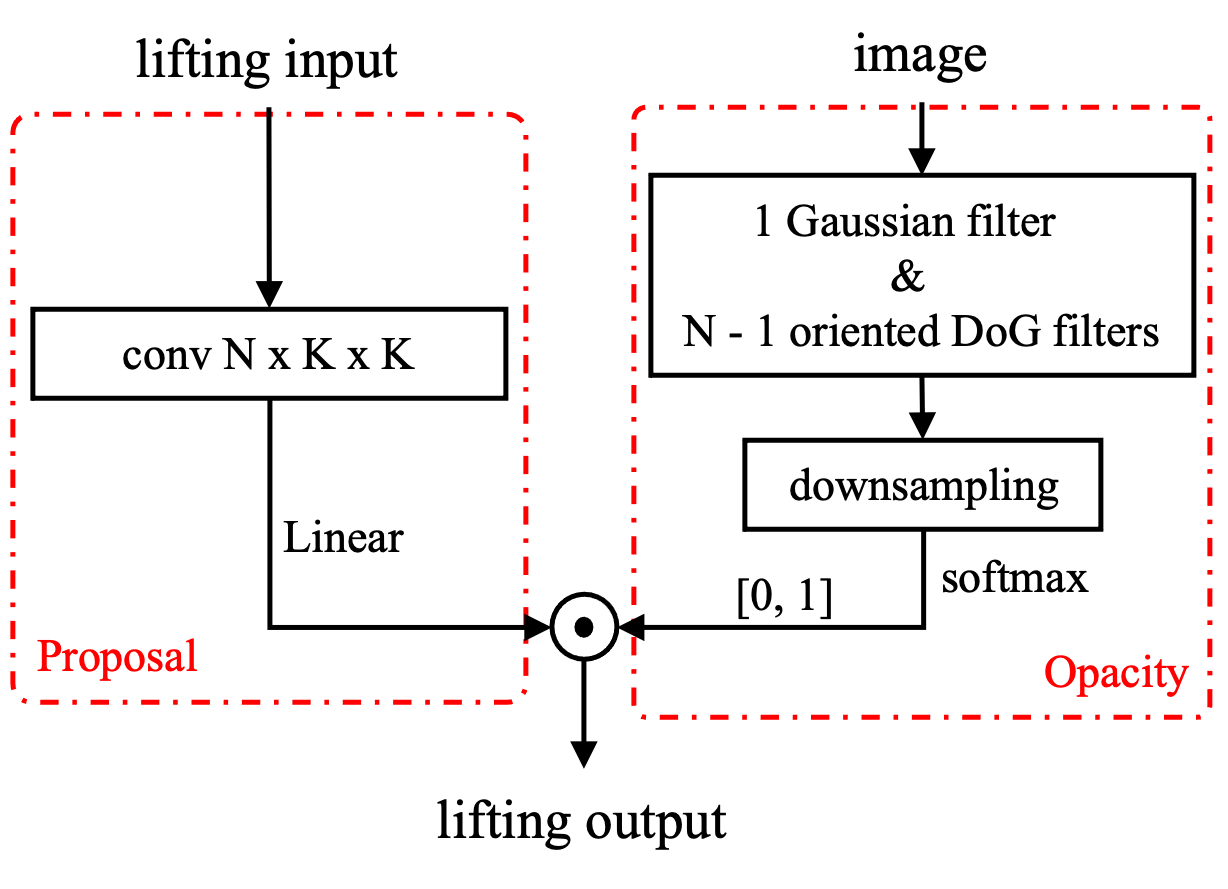}
	\caption{The proposed method to create oracle opacities. The notation \emph{conv} $N$ x $K$ x $K$ represents the convolutional layer with $N$ channels (or filters) and kernel support $K \text{~x~} K$.}
	\label{fig:oracle_opacities}
\end{figure}

In this paper, the heuristics that we employ to derive the oracle opacities are a non-oriented Gaussian filter and a collection of oriented derivative of Gaussian (DoG) filters with standard deviation $\sigma_d$, as shown in Fig.~\ref{fig:oracle_opacities}.
We let $\sigma_d$ increase according to the scale of each decomposition level $d$; that is $\sigma_d = 2^{d-1}, d = 1, 2, 3, \cdots$, where $d = 1$ denotes the finest level of decomposition.
The non-oriented Gaussian filter always corresponds to the frozen proposal in \emph{Stage 1}, because it is reasonable to utilize the fixed lifting filter when there is no specific orientation.
The rationale behind this orientation-based approach rests in the importance of local geometric flow for eliminating redundancy in the wavelet transform, as discussed in Section~\ref{subsec:significance_proposal_opacity_topology}.

As we have explained, these data-dependent opacities used in \emph{Stage 1} can be understood as oracle values that are unavailable to the decoder during reconstruction.
However, surprisingly the true opacity network used in \emph{Stage 2} turns out to be able to discover opacities whose utility is comparable with, and even slightly superior to the utility of oracle opacities, as we shall see in Section~\ref{subsubsec:significance_pre_training}.
This indicates that the heuristic model behind the oracle opacities is sensible.

\subsubsection{Pre-training with progressive selection}
\label{subsubsec:progressive_selection}
Although the oracle-opacity pre-training schedule is applicable to the proposal-opacity network topology, it cannot be generalized well to other network architectures, such as \emph{iWave} and \emph{iWave++} topologies investigated in this paper.

Therefore, we also consider another pre-training strategy with a \emph{progressive selection} objective as proposed in \cite{ma2019iwave}; that is
\begin{align}
J_{ps} = \sum^{3d+1}_{i=1} E \left[ \left \| \mathbf{x} - \hat{\mathbf{x}}_i \right \|^2_2 \right],
\label{eq:J_ps}
\end{align}
where $\hat{\mathbf{x}}_i$ represents the reconstructed image obtained using only the first $i$ transformed subbands, without any quantization; $i$ starts with the 
In this way, the composite objective $J_{ps}$ can be understood as encouraging learned lifting operators to compact the information of the input image $\mathbf{x}$ into as few subbands as possible; this is generally desirable for image compression.
This pre-training strategy with progressive selection is applicable to a wide variety of network topologies, providing a fair basis for comparisons between different network topologies in this paper.

\section{Experimental Results}
\label{sec:experimental_results}

In this section, we first empirically demonstrate the effectiveness of the pre-training strategies proposed in Section~\ref{subsec:pre_training}.
Subsequently, we study the value brought by three different network topologies mentioned in Section~\ref{sec:proposal_opacity_topology}.
Furthermore, we develop a sequence of experiments to study different aspects (depth, diversity and region of support) of the investigated lifting structures listed in Section~\ref{sec:investigated_lifting_structure}.
Ultimately, these experiments give us guidance on how to strategically deploy neural networks to enhance the base wavelet transform for compression, balancing coding performance with computational complexity and region of support. 
Note that the source code of this paper, along with all the training and testing datasets, are available on GitHub\footnote{https://github.com/xinyue-li3/learned-wavelet-like-transforms}.

\subsection{Experimental Settings}
\label{subsec:experimental_settings}

\subsubsection{Training Phase}
In this paper, Keras with TensorFlow backend and the Adam algorithm \cite{kingma2014adam}
are employed for training, with 75 image batches comprising 16 patches of size 256 x 256 from the DIV2K image dataset\footnote{https://data.vision.ee.ethz.ch/cvl/DIV2K/}.
We employ 5 levels of the wavelet decomposition during training, and aim to jointly train only one set of learned lifting steps, which can be applied to all levels of decomposition, as well as a wide range of bit-rates.
This goal is explicitly chosen for the reasons elaborated in Section~\ref{subsec:proposed_lifting_networks}.
We employ the JPEG 2000 developer toolkit, Kakadu software\footnote{https://kakadusoftware.com}, to encode and decode the coefficients of all learned wavelet-like transforms, during both training and inference.
It is important to highlight here that we encode the image only once to high quality; all reduced-quality representations of the image are derived from this one bit-stream by discarding irrelevant bits, benefiting from the quality scalability feature of JPEG~2000.
This is a remarkable difference between our work in this paper and most of other existing compression schemes.

\subsubsection{Testing Phase}
In this paper, four categorized datasets are used during testing, in order to demonstrate the merits of different lifting structures in various scenarios.
Note that none of these images are used during training.

\textbf{Category 1}: All images within this class have highly structured features, i.e. edges are either consistently oriented or significantly distinct from background textures. Two datasets are included in this class: a) Tecnick Sampling Dataset\footnote{https://testimages.org/}, from which 20 images are chosen with size 480 x 480; b) DIV2K Dataset, from which 30 images are chosen with size 1024 x 2048. We name these two dataset as \emph{Tecnick-Cat1} and \emph{DIV2K-Cat1}, respectively.

\textbf{Category 2}: All images in this category come with reasonably clear edges, while background textures are more complicated than those in Category 1. The dataset employed in this class is DIV2K dataset, from which another 70 images of size 1024 x 2048 are chosen; it is denoted by \emph{DIV2K-Cat2}.

\textbf{Category 3}: All images in this category are considered to be ``hard-to-code'', with one or more following properties: nearly no clear orientations; majority of the image is excessively blurred; and/or most orientations are horizontal or vertical, which are well handled by the wavelet transform. The dataset employed is Challenges on Learned Image Compression 2019 test set\footnote{http://clic.compression.cc/2019/challenge/}, from which 15 images of size 1024 x 2048 are chosen; this is denoted as \emph{CLIC2019-Cat3}.

\subsection{Evaluation Metrics}
\label{subsec:evaluation_metrics}
We consider evaluating the performance of all lifting structures both quantitatively and qualitatively.
In terms of quantitative measurements, three widely used metrics are employed -- Peak Signal-to-Noise Ratio (PSNR), Structural Similarity (SSIM), Multi-Scale Structural Similarity (MS-SSIM). All these metrics are measured and averaged for each dataset over the range of bit-rates from $0.1$bpp to $1.0$bpp, from which Bj{\o}ntegaard (BD) rate savings (in \%) are obtained \cite{bjontegaard2001calculation}.
With regard to qualitative assessment, we provide examples for the LL bands produced after analysis, and the full reconstructed images produced after synthesis, by different lifting structures in this paper. 

\subsection{Results and Discussion}
\label{subsec:results_and_discussions}

\subsubsection{Significance of the investigated pre-training strategies}
\label{subsubsec:significance_pre_training}
We first empirically study the significance of the two investigated pre-training schedules in Section~\ref{subsec:pre_training}.
For the sake of simplicity, we focus only on learning the predict-update lifting structure shown in Fig.~\ref{fig:predict_update}(b), along with the proposal-opacity network topology for all learned lifting operators in this section. 
Further studies on other investigated lifting structures and network topologies are provided shortly.

\begin{figure}[htb]
	\centering
	\subfloat[]{
	\includegraphics[width=1.0\linewidth]{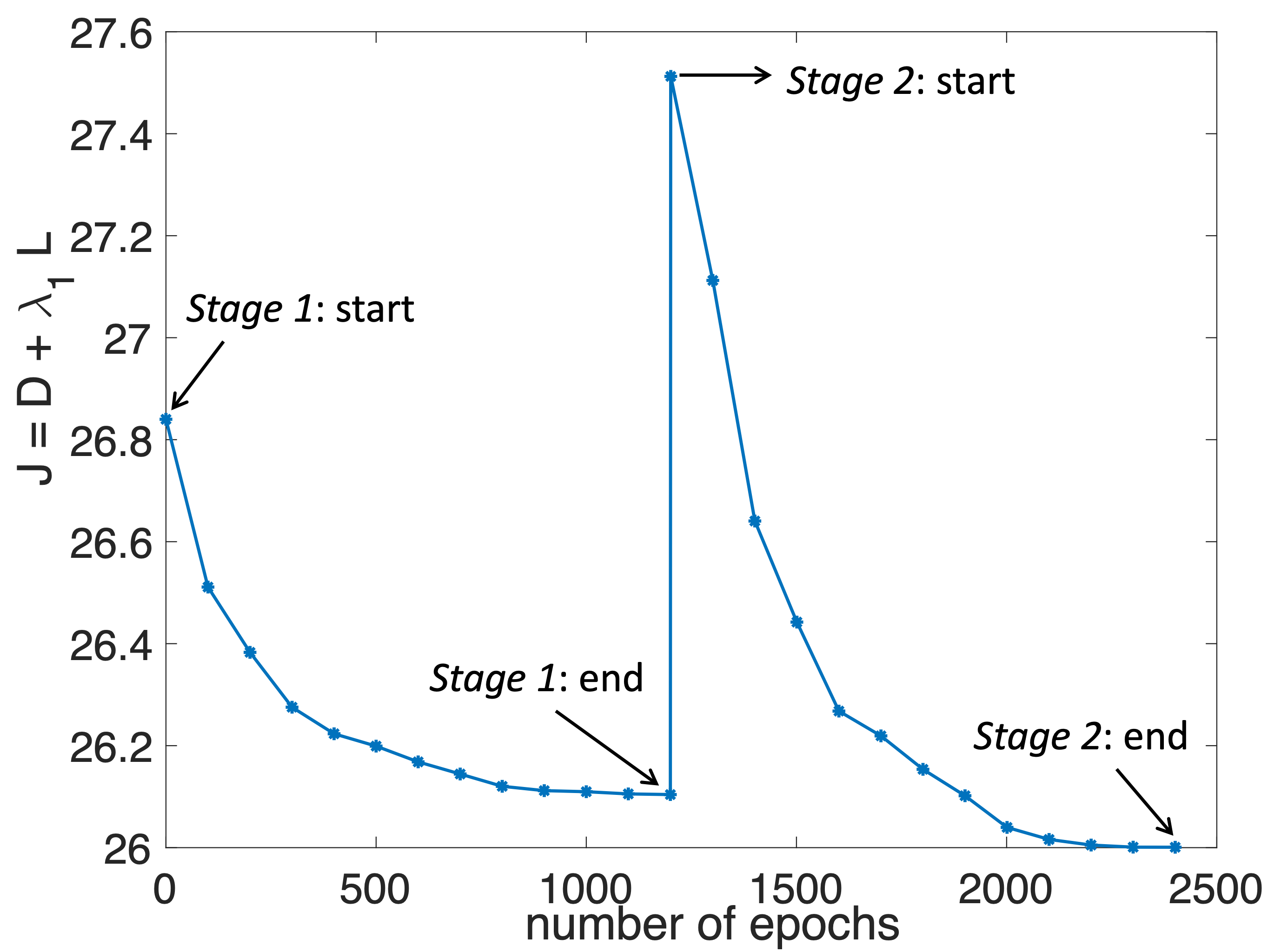}}\\
	\subfloat[]{
	\includegraphics[width=1.0\linewidth]{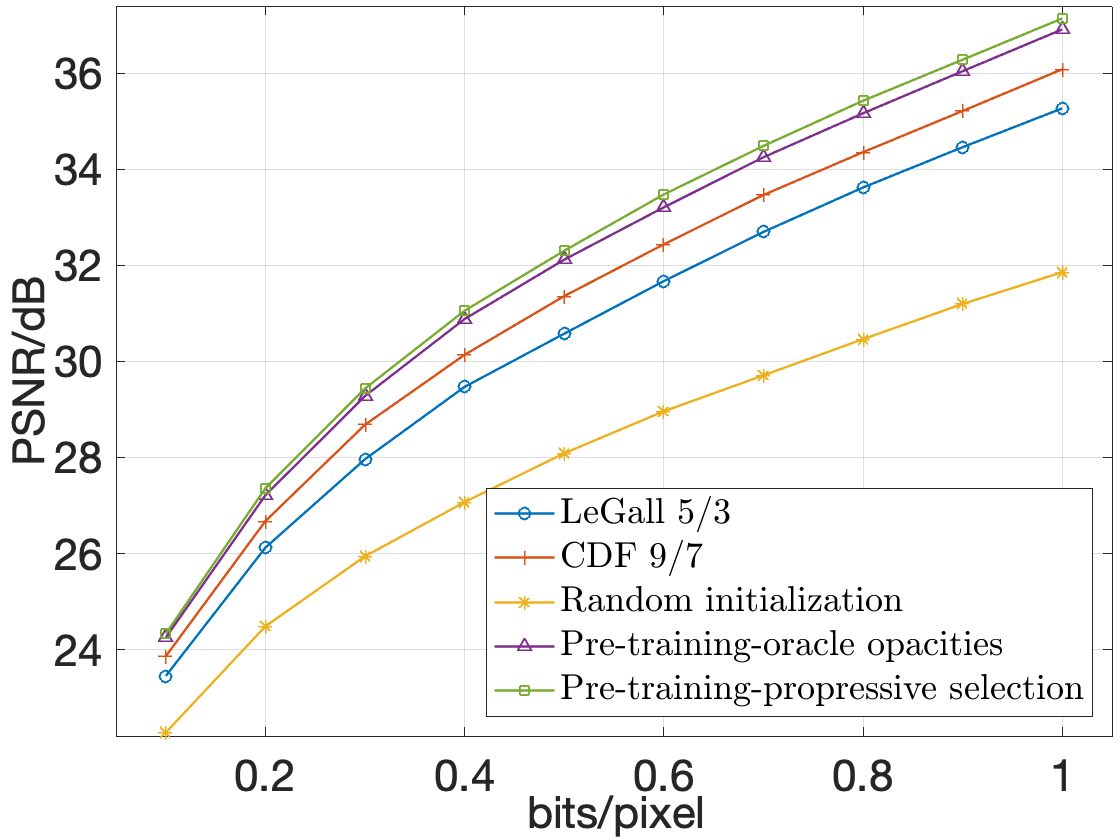}}
	\caption{(a) The envelope of the rate-distortion objective in \eqref{eq:R_D} throughout \emph{Stage 1} and \emph{Stage 2} of the proposed training schedule. (b) Comparisons of the average PSNR improvements for \emph{Tecknick-Cat1} dataset using different pre-training strategies. Here we focus only on learning the predict-update lifting structure shown in Fig.~\ref{fig:predict_update}(b), along with the proposal-opacity network topology for all learned lifting operators.}
	\label{fig:pre_training_strategy}
\end{figure}

As shown in Fig.~\ref{fig:pre_training_strategy}(b), if we adopt only random initialisation and without any proper pre-training, the learned lifting operators within the base wavelet transform can wind up exhibiting much worse coding performance than the conventional wavelet transforms across all four datasets for average PSNR over the range of bit-rates from $0.1$bpp to $1.0$bpp.
This observation reinforces the need for a robust pre-training strategy.

As we have discussed in Section~\ref{subsubsec:oracle_training_schedule}, the oracle-opacity pre-training schedule relies on heuristics to produce oracle values, which are expected to be useful and potentially reproducible by the true opacity network in the inference machine.
The reproducibility needs confirmation, since the oracle opacities are derived directly from input images, which are generally not available during inference.  

To address this, we plot the envelope of the rate-distortion objective in \eqref{eq:R_D} throughout \emph{Stage 1} and \emph{Stage 2} of the proposed pre-training schedule.
Fig.~\ref{fig:pre_training_strategy}(a) demonstrates that the true opacity branches are indeed capable of discovering opacities whose utility is comparable with and even superior to the utility of oracle opacities, producing a slightly lower rate-distortion value at the end of \emph{Stage 2} compared with that of \emph{Stage 1}.
This observation strongly reinforces the appropriateness of the oracle-opacity pre-training schedule.

As it turns out in Fig.~\ref{fig:pre_training_strategy}(b), pre-training with progressive selection outperforms random initialisation by a large margin.
More importantly, the oracle-opacity pre-training schedule results in similar performance as the one with progressive selection, despite being two very different approaches.
Although we only show results for \emph{Tecknick-Cat1} dataset evaluated using PSNR, the conclusion stays the same for other test datasets evaluated using SSIM and MS-SSIM.
This gives us more confidence that the two investigated pre-training strategies converge to good solutions.

\subsubsection{Value brought by different network topologies and lifting structures}
\label{subsubsec:values_NN_topology}
In this section, we empirically study a list of variations to understand the value brought by different network topologies, as well as the impact of different lifting structures.
These variations are referred to as \emph{PO-P-U}, \emph{iWave-P-U}, \emph{iWave++-P-U}, \emph{PO-U-P}, \emph{iWave-U-P} and 
\emph{PO-Hybrid(9/7)}.
We employ prefixes \emph{PO-}, \emph{iWave-} and \emph{iWave++} to denote the proposal-opacity, iWave\cite{ma2019iwave} and iWave++\cite{ma2020end} network topologies.
We use suffixes \emph{-P-U}, \emph{-U-P} and \emph{Hybrid} to represent the predict-update, update-predict and hybrid lifting structures, as seen in Section~\ref{subsec:predict-update}, Section~\ref{subsec:update-predict} and Section~\ref{subsec:hybrid_structure}.
Here \emph{(9/7)} denotes the CDF 9/7 base wavelet transform.

For fair comparison, we pre-train all these variations using progressive selection (see Section~\ref{subsubsec:progressive_selection}), followed by the end-to-end optimisation (see Section~\ref{subsec:end2end_learning_strategy}).
Although we only show results for \emph{Tecknick-Cat1} dataset evaluated using PSNR, the conclusion stays the same for other test datasets evaluated using SSIM and MS-SSIM.

\begin{figure}[htb]
	\centering
	\includegraphics[width=1.0\linewidth]{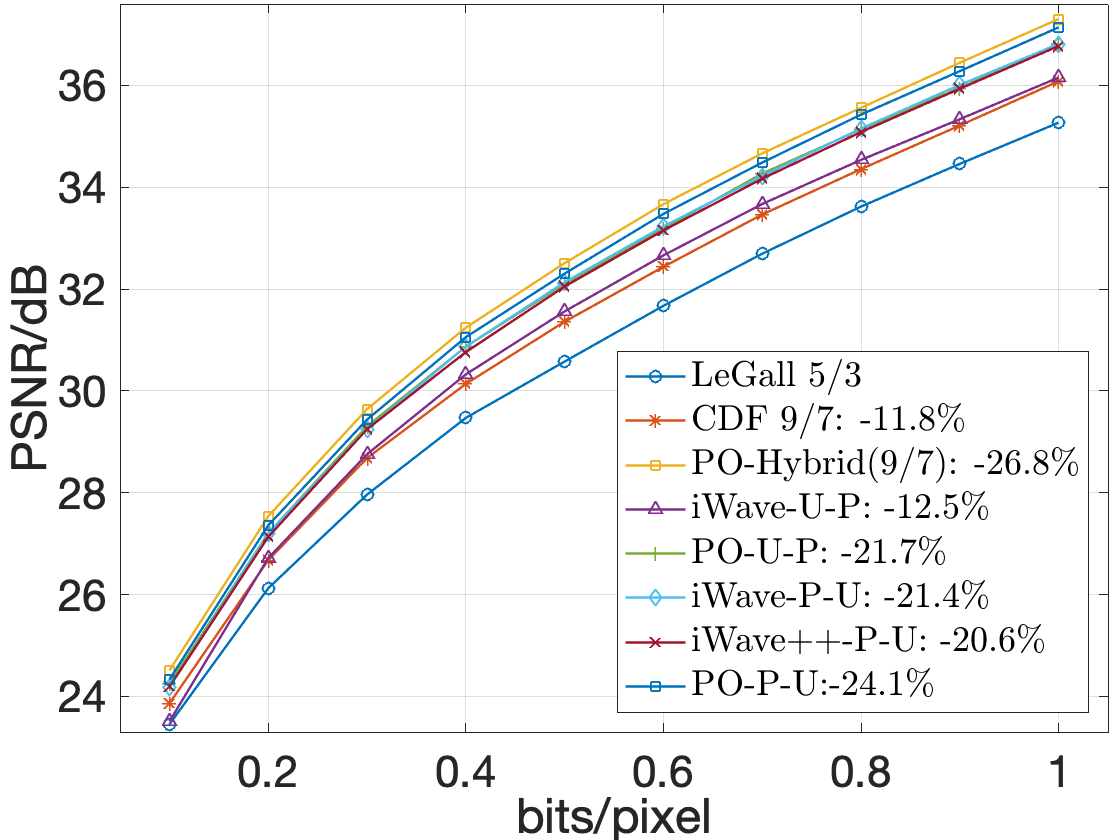}
	\caption{The rate-distortion performance of the investigated variations on \emph{Tecnick-Cat1} dataset, evaluated by PSNR in dB. Bj{\o}ntegaard (BD) rate savings (in \%) are displayed next to legends; we use the LeGall 5/3 wavelet transform as the anchor here.}
	\label{fig:values_topologies}
\end{figure}

Fig.~\ref{fig:values_topologies} strongly suggests that the proposal-opacity network topology is better than the iWave and iWave++ topologies, achieving 24.1\% averaged bit-rate savings over the LeGall 5/3 wavelet transform, when employed in the predict-update lifting structure.
In addition, we observe that the predict-update lifting structure outperforms the update-predict structure by a large margin, regardless of which network topology is used.
This is in contrast to the conclusion made in \cite{ma2019iwave}; the reason is that we employ a full end-to-end optimization framework for training, while \cite{ma2019iwave} only trains the learned lifting operators using the pre-training with progressive selection seen in Section~\ref{subsubsec:progressive_selection}.

More importantly, we see that the hybrid lifting structure performs best amongst all investigated structures, achieving 26.8\% averaged bit-rate savings over the LeGall 5/3 wavelet transform, when employing the proposal-opacity networks.

\subsubsection{Impact of increasing the depth of lifting structures}
\label{subsubsec:merits_depth}
Although the hybrid lifting structure performs better than the predict-update structure shown in Section~\ref{subsec:predict-update}, it essentially comes with more lifting steps -- fixed lifting steps in the base wavelet transform plus two learned lifting steps.
Therefore, the question remains if the conclusion in Section~\ref{subsubsec:values_NN_topology} still holds, when we employ more learned lifting steps for the predict-update structure.

To answer this question, we employ the proposal-opacity network topology for all learned lifting operators in this section, and examine the performance of the following variations: \emph{Hybrid(5/3)-5c}, \emph{Hybrid(9/7)-5c}, \emph{Custom-4S-5c}, \emph{Custom-4MS-5c} and \emph{Custom-5S-5c}.
\emph{Hybrid(5/3)} and \emph{Hybrid(9/7)} denote the hybrid structure with only two learned lifting steps shown in Fig.~\ref{fig:hybrid_LS}, using LeGall 5/3 and CDF 9/7 wavelet transforms as the base wavelet transform respectively.
\emph{Custom-4S} denotes the predict-update  structure with four learned lifting steps shown in Fig.~\ref{fig:predict_update}(b).
\emph{Custom-4MS} denotes the predict-update  structure with four modified learned lifting steps, by replacing the last update step $\mathcal{U}^H$ in Fig.~\ref{fig:predict_update}(b) with the high-to-low step $\mathcal{T}^A_{H2L}$.
\emph{Custom-5S} denotes the predict-update  structure with five learned lifting steps shown in Fig.~\ref{fig:Custom_5S}.
We use \emph{-5c} to indicate the number of channels $N = 5$ used in the proposal-opacity network topology, as shown in Fig.~\ref{fig:proposed_lifting_networks} and Fig.~\ref{fig:proposed_H2L_L2H_networks}.

It is worthwhile to highlight here that each learned lifting operator comes with a substantially large region of support as well as high computational complexity.
Therefore, it is important to examine whether any benefits in coding performance can be justified in light of the negative impacts on complexity and region of support, by adding more learned lifting steps.
\begin{table}[tb!]
	\centering
	\caption{Impact of the number of learned lifting steps, the number of channels for each learned lifting operator and spatial support of each learned lifting operator on coding efficiency. The table shows BD-rate improvements for average PSNR, SSIM and MS-SSIM metrics over the LeGall 5/3 and the CDF 9/7 wavelet transform. Results are obtained with bit-rates between $0.1$bpp to $1.0$bpp.}
	\resizebox{0.5\textwidth}{!}{
	\begin{tabular}{ll*{7}{S[table-format=-2.1]}}
		\toprule
		& & \multicolumn{7}{c}{BD-rate for PSNR}\\
		\cmidrule(lr){3-9}
		& & 
		\multicolumn{1}{c}{\makecell{Hybrid\\(5/3)-5c}} & \multicolumn{1}{c}{\makecell{Hybrid\\(9/7)-5c}} & \multicolumn{1}{c}{\makecell{Custom\\-4S-5c}} & \multicolumn{1}{c}{\makecell{Custom\\-4MS-5c}} & \multicolumn{1}{c}{\makecell{Custom\\-5S-5c}} & \multicolumn{1}{c}{\makecell{Hybrid\\(9/7)-9c}} & \multicolumn{1}{c}{\makecell{Hybrid\\(9/7)-9c\\-compact}} \\
		\cmidrule{2-9}
		\multirow{4}{*}{\rotatebox{90}{LeGall5/3}} 
		& Tecknick-Cat1 & -17.4\% & -21.9\% & -17.3\% & -18.4\% & -20.3\% & -24.0\% & -25.6\% \\
		& DIV2K-Cat1 & -14.4\% & -19.6\% & -17.1\% & -18.3\% & -19.7\% & -22.2\% & -24.0\% \\
		& DIV2K-Cat2 & -12.5\% & -15.9\% & -13.4\% & -14.2\% & -15.6\% & -17.9\% & -19.4\% \\
		& CLIC2019-Cat3 & -7.3\% & -10.8\% & -8.2\% & -8.7\% & -9.8\% & -12.2\% & -12.8\% \\
		\cmidrule{2-9}
		\multirow{4}{*}{\rotatebox{90}{CDF9/7}}    
		& Tecknick-Cat1 & -6.5\% & -11.4\% & -6.3\% & -7.5\% & -9.5\% & -13.7\% & -15.5\% \\
		& DIV2K-Cat1 & -3.8\% & -9.7\% & -7.0\% & -8.2\% & -9.9\% & -12.6\% & -14.6\% \\
		& DIV2K-Cat2 & -3.8\% & -7.5\% & -4.9\% & -5.7\% & -7.2\% & -9.7\% & -11.4\% \\
		& CLIC2019-Cat3 & -0.5\% & -4.2\% & -1.5\% & -1.8\% & -3.2\% & -5.7\% & -6.4\% \\
		\bottomrule
	\end{tabular}}\\
	\resizebox{0.5\textwidth}{!}{
	\begin{tabular}{ll*{7}{S[table-format=-2.1]}}
		\toprule
		& & \multicolumn{7}{c}{BD-rate for SSIM}\\
		\cmidrule(lr){3-9}
		& & 
		\multicolumn{1}{c}{\makecell{Hybrid\\(5/3)-5c}} & \multicolumn{1}{c}{\makecell{Hybrid\\(9/7)-5c}} & \multicolumn{1}{c}{\makecell{Custom\\-4S-5c}} & \multicolumn{1}{c}{\makecell{Custom\\-4MS-5c}} & \multicolumn{1}{c}{\makecell{Custom\\-5S-5c}} & \multicolumn{1}{c}{\makecell{Hybrid\\(9/7)-9c}} & \multicolumn{1}{c}{\makecell{Hybrid\\(9/7)-9c\\-compact}} \\
		\cmidrule{2-9}
		\multirow{4}{*}{\rotatebox{90}{LeGall5/3}} 
		& Tecknick-Cat1 & -15.5\% & -16.9\% & -15.8\% & -16.7\% & -18.6\% & -19.9\% & -22.5\% \\
		& DIV2K-Cat1 & -13.8\% & -15.7\% & -15.1\% & -16.4\% & -18.4\% & -18.9\% & -21.5\% \\
		& DIV2K-Cat2 & -12.8\% & -13.8\% & -13.6\% & -14.3\% & -16.1\% & -16.4\% & -18.3\% \\
		& CLIC2019-Cat3 & -7.5\% & -8.2\% & -7.8\% & -8.1\% &
-9.5\% & -10.1\% & -10.9\% \\
		\cmidrule{2-9}
		\multirow{4}{*}{\rotatebox{90}{CDF9/7}}    
		& Tecknick-Cat1 & -10.0\% & -11.5\% & -10.2\% & -11.1\% & -13.2\% & -14.5\% & -17.3\% \\
		& DIV2K-Cat1 & -8.2\% & -10.1\% & -9.5\% & -10.9\% & -12.9\% & -13.5\% & -16.2\% \\
		& DIV2K-Cat2  & -6.3\% & -7.5\% & -7.2\% & -7.9\% & -9.8\% & -10.1\% & -12.2\% \\
		& CLIC2019-Cat3 & -3.3\% & -4.1\% & -3.6\% & -4.0\% & -5.4\% & -6.0\% & -6.9\% \\
		\bottomrule
	\end{tabular}}\\
	\resizebox{0.5\textwidth}{!}{
	\begin{tabular}{ll*{7}{S[table-format=-2.1]}}
		\toprule
		& & \multicolumn{7}{c}{BD-rate for MS-SSIM}\\
		\cmidrule(lr){3-9}
		& & 
		\multicolumn{1}{c}{\makecell{Hybrid\\(5/3)-5c}} & \multicolumn{1}{c}{\makecell{Hybrid\\(9/7)-5c}} & \multicolumn{1}{c}{\makecell{Custom\\-4S-5c}} & \multicolumn{1}{c}{\makecell{Custom\\-4MS-5c}} & \multicolumn{1}{c}{\makecell{Custom\\-5S-5c}} & \multicolumn{1}{c}{\makecell{Hybrid\\(9/7)-9c}} & \multicolumn{1}{c}{\makecell{Hybrid\\(9/7)-9c\\-compact}} \\
		\cmidrule{2-9}
		\multirow{4}{*}{\rotatebox{90}{LeGall5/3}} 
		& Tecknick-Cat1 & -13.6\% & -17.2\% & -15.1\% & -16.3\% & -18.8\% & -19.5\% & -21.9\% \\
		& DIV2K-Cat1 & -13.1\% & -16.8\% & -14.5\% & -15.9\% & -18.5\% & -19.0\% & -21.5\% \\
		& DIV2K-Cat2 & -12.7\% & -15.0\% & -13.8\% & -14.6\% & -17.2\% & -17.1\% & -18.8\% \\
		& CLIC2019-Cat3 & -8.9\% & -11.0\% & -8.9\% & -9.5\% & -12.5\% & -12.3\% & -13.1\% \\
		\cmidrule{2-9}
		\multirow{4}{*}{\rotatebox{90}{CDF9/7}}    
		& Tecknick-Cat1 & -5.1\% & -9.0\% & -6.7\% & -8.0\% & -10.6\% & -11.5\% & -14.1\% \\
		& DIV2K-Cat1 & -9.9\% & -7.5\% & -5.1\% & -6.6\% & -9.5\% & -10.0\% & -12.7\% \\
		& DIV2K-Cat2 & -3.4\% & -5.9\% & -4.7\% & -5.5\% & -8.3\% & -8.2\% & -10.0\% \\
		& CLIC2019-Cat3 & 0.6\% & -1.7\% & 0.5\% & -0.2\% & -3.4\% & -3.1\% & -4.1\% \\
		\bottomrule
	\end{tabular}}
	\label{table:merits_depth}
\end{table}

\begin{figure}[tb]
	\centering
	\includegraphics[width=1.0\linewidth]{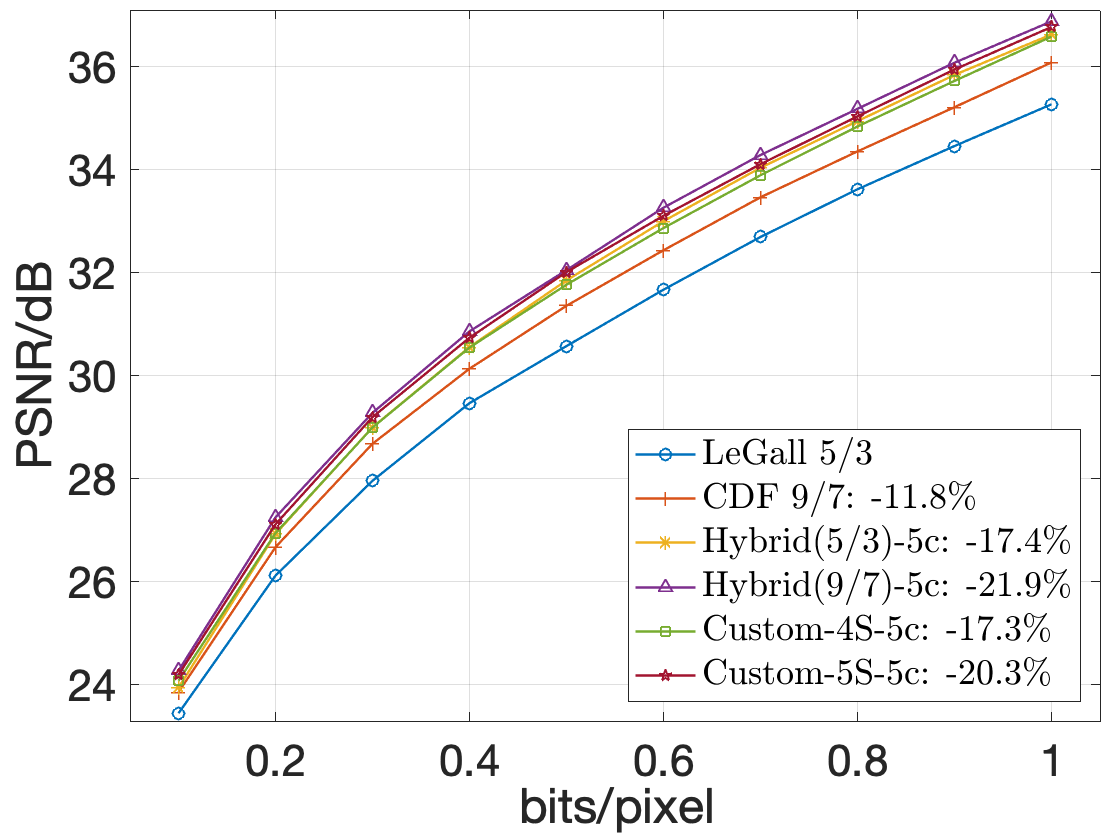}
	\caption{Comparisons of the average PSNR improvements over the LeGall 5/3 and CDF 9/7 wavelet transforms across \emph{Tecnick-Cat1} dataset, using lifting structures with different numbers of learned lifting steps. Bj{\o}ntegaard (BD) rate savings (in \%) are displayed next to legends; we use the LeGall 5/3 wavelet transform as the anchor here.}
	\label{fig:merits_depth}
\end{figure}

Since all the variations investigated in this section involve the proposal-opacity network topology, we apply the oracle-opacity pre-training schedule universally here.
The BD-rate savings (in \%) for average PSNR, SSIM and MS-SSIM over the range of bit-rates from $0.1$bpp to $1.0$bpp across all four datasets are provided in Table~\ref{table:merits_depth}.
Key rate-distortion curves are shown in Fig.~\ref{fig:merits_depth}.

\begin{figure}[tb!]
	\centering
	\begin{center}
		\begin{tikzpicture}
		\node(a){\subfloat[The original image, cropped from image 9 of Tecnick-Cat1 dataset]{%
				\includegraphics[width=0.5\linewidth, trim={0cm 3cm 11cm 9cm},clip]{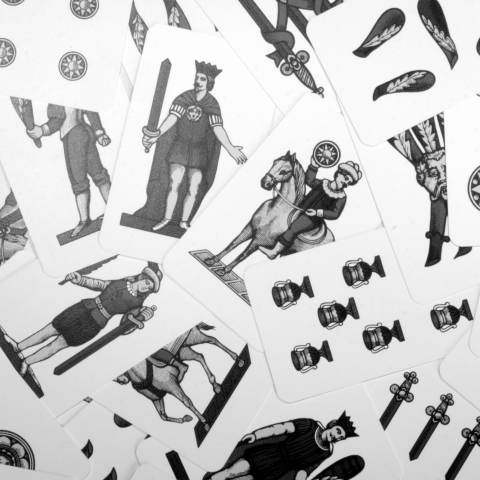}}};
		\end{tikzpicture}%
		\begin{tikzpicture}
		\node(a){\subfloat[LeGall 5/3, PSNR=26.97dB, SSIM=0.862, MS-SSIM=0.9694]{%
				\includegraphics[width=0.5\linewidth, trim={0cm 3cm 11cm 9cm},clip]{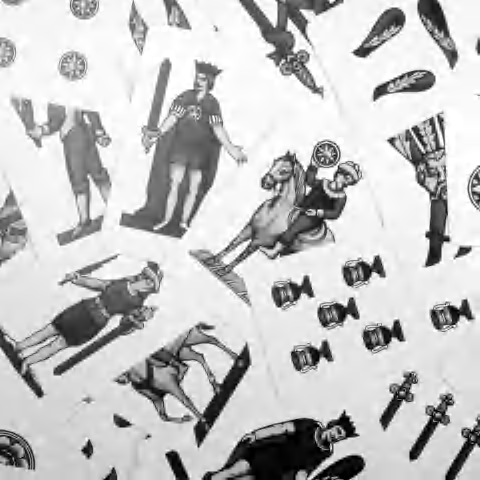}}};
		\node at(a.center)[draw, red,line width=0.5pt,ellipse, minimum width=35pt, minimum height=20pt,xshift=5pt,yshift=40pt,rotate=0]{};
		\node at(a.center)[draw, red,line width=0.5pt,rectangle, minimum width=30pt, minimum height=30pt,xshift=45pt,yshift=40pt,rotate=0]{};
		\node at(a.center)[draw, red,line width=0.5pt,rectangle, minimum width=45pt, minimum height=45pt,xshift=-35pt,yshift=-10pt,rotate=0]{};
		\end{tikzpicture}\\
		\begin{tikzpicture}
		\node(a){\subfloat[CDF 9/7, PSNR=$27.36$dB, SSIM=$0.854$, MS-SSIM=$0.9681$]{%
			\includegraphics[width=0.5\linewidth, trim={0cm 3cm 11cm 9cm},clip]{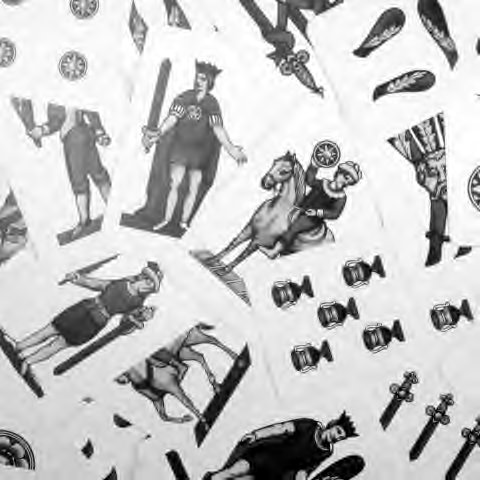}}};
			\node at(a.center)[draw, red,line width=0.5pt,ellipse, minimum width=35pt, minimum height=20pt,xshift=5pt,yshift=40pt,rotate=0]{};
		\node at(a.center)[draw, red,line width=0.5pt,rectangle, minimum width=30pt, minimum height=30pt,xshift=45pt,yshift=40pt,rotate=0]{};
		\node at(a.center)[draw, red,line width=0.5pt,rectangle, minimum width=45pt, minimum height=45pt,xshift=-35pt,yshift=-10pt,rotate=0]{};
		\end{tikzpicture}%
		\begin{tikzpicture}
		\node(a){\subfloat[Hybrid(5/3)-5c, PSNR=$28.19$dB, SSIM=$0.875$, MS-SSIM=$0.972$]{%
		\includegraphics[width=0.5\linewidth, trim={0cm 3cm 11cm 9cm},clip]{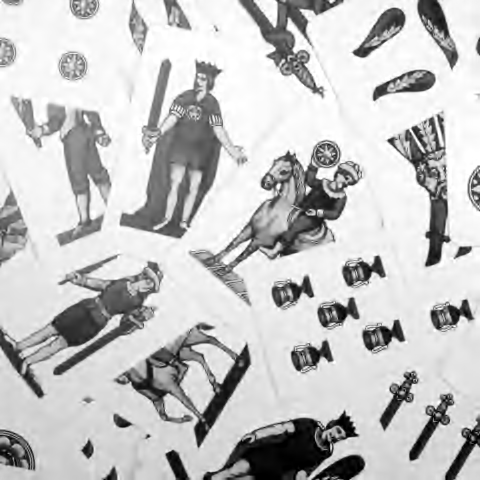}}};
		\node at(a.center)[draw, red,line width=0.5pt,ellipse, minimum width=35pt, minimum height=20pt,xshift=5pt,yshift=40pt,rotate=0]{};
		\node at(a.center)[draw, red,line width=0.5pt,rectangle, minimum width=30pt, minimum height=30pt,xshift=45pt,yshift=40pt,rotate=0]{};
		\node at(a.center)[draw, red,line width=0.5pt,rectangle, minimum width=45pt, minimum height=45pt,xshift=-35pt,yshift=-10pt,rotate=0]{};
		\end{tikzpicture}\\
		\begin{tikzpicture}
		\node(a){\subfloat[Hybrid(9/7)-5c, PSNR=$28.41$dB, SSIM=$0.870$, MS-SSIM=$0.971$]{%
				\includegraphics[width=0.5\linewidth, trim={0cm 3cm 11cm 9cm},clip]{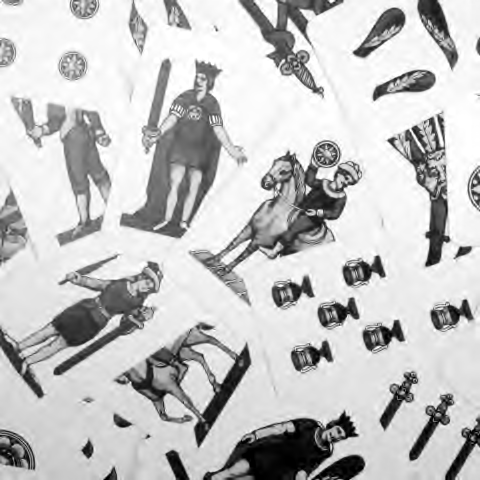}}};
				\node at(a.center)[draw, red,line width=0.5pt,ellipse, minimum width=35pt, minimum height=20pt,xshift=5pt,yshift=40pt,rotate=0]{};
		\node at(a.center)[draw, red,line width=0.5pt,rectangle, minimum width=30pt, minimum height=30pt,xshift=45pt,yshift=40pt,rotate=0]{};
		\node at(a.center)[draw, red,line width=0.5pt,rectangle, minimum width=45pt, minimum height=45pt,xshift=-35pt,yshift=-10pt,rotate=0]{};
		\end{tikzpicture}%
		\begin{tikzpicture}
		\node(a){\subfloat[Custom-5S-5c, PSNR=$28.31$dB, SSIM=$0.894$, MS-SSIM=$0.975$]{%
				\includegraphics[width=0.5\linewidth, trim={0cm 3cm 11cm 9cm},clip]{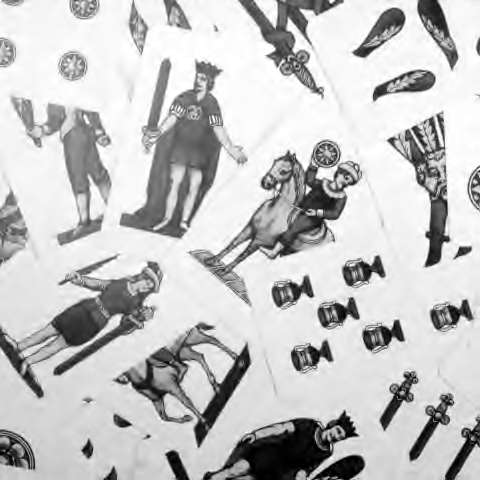}}};
				\node at(a.center)[draw, red,line width=0.5pt,ellipse, minimum width=35pt, minimum height=20pt,xshift=5pt,yshift=40pt,rotate=0]{};
		\node at(a.center)[draw, red,line width=0.5pt,rectangle, minimum width=30pt, minimum height=30pt,xshift=45pt,yshift=40pt,rotate=0]{};
		\node at(a.center)[draw, red,line width=0.5pt,rectangle, minimum width=45pt, minimum height=45pt,xshift=-35pt,yshift=-10pt,rotate=0]{};
		\end{tikzpicture}\\
		\begin{tikzpicture}
		\node(a){\subfloat[Hybrid(9/7)-9c, PSNR=$28.8$dB, SSIM=$0.898$, MS-SSIM=$0.973$]{%
				\includegraphics[width=0.5\linewidth, trim={0cm 3cm 11cm 9cm},clip]{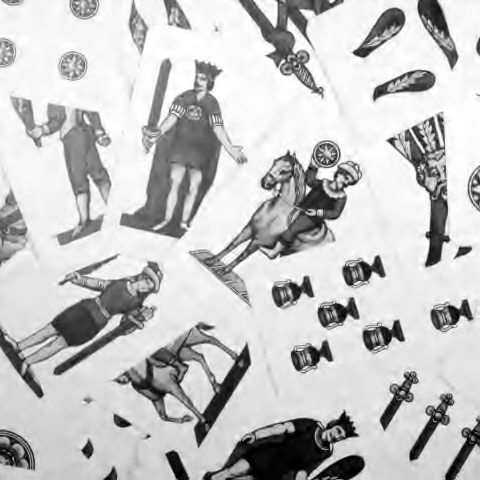}}};
				\node at(a.center)[draw, red,line width=0.5pt,ellipse, minimum width=35pt, minimum height=20pt,xshift=5pt,yshift=40pt,rotate=0]{};
		\node at(a.center)[draw, red,line width=0.5pt,rectangle, minimum width=30pt, minimum height=30pt,xshift=45pt,yshift=40pt,rotate=0]{};
		\node at(a.center)[draw, red,line width=0.5pt,rectangle, minimum width=45pt, minimum height=45pt,xshift=-35pt,yshift=-10pt,rotate=0]{};
		\end{tikzpicture}%
		\begin{tikzpicture}
		\node(a){\subfloat[Hybrid(9/7)-9c-compact,PSNR= 29.1dB,SSIM=$0.891$,MSSSIM=$0.973$]{%
				\includegraphics[width=0.5\linewidth, trim={0cm 3cm 11cm 9cm},clip]{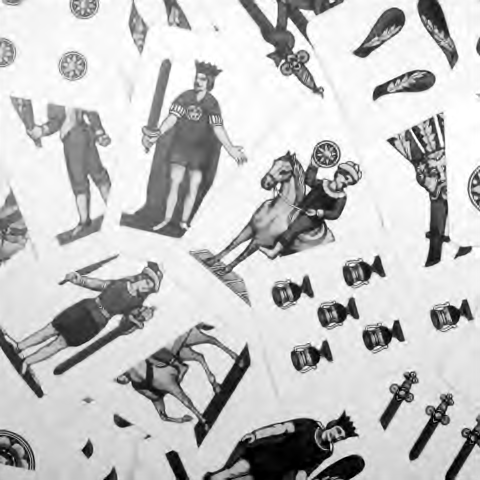}}};
				\node at(a.center)[draw, red,line width=0.5pt,ellipse, minimum width=35pt, minimum height=20pt,xshift=5pt,yshift=40pt,rotate=0]{};
		\node at(a.center)[draw, red,line width=0.5pt,rectangle, minimum width=30pt, minimum height=30pt,xshift=45pt,yshift=40pt,rotate=0]{};
		\node at(a.center)[draw, red,line width=0.5pt,rectangle, minimum width=45pt, minimum height=45pt,xshift=-35pt,yshift=-10pt,rotate=0]{};
		\end{tikzpicture}
	\end{center}
	\caption{Examples of fully reconstructed images, obtained from different lifting structures with various numbers of learned lifting steps, diversities and regions of support. We compare these results with two conventional wavelet transforms: the LeGall 5/3 and the CDF 9/7 wavelet transforms.}
	\label{fig:reconstructed_img}
\end{figure}
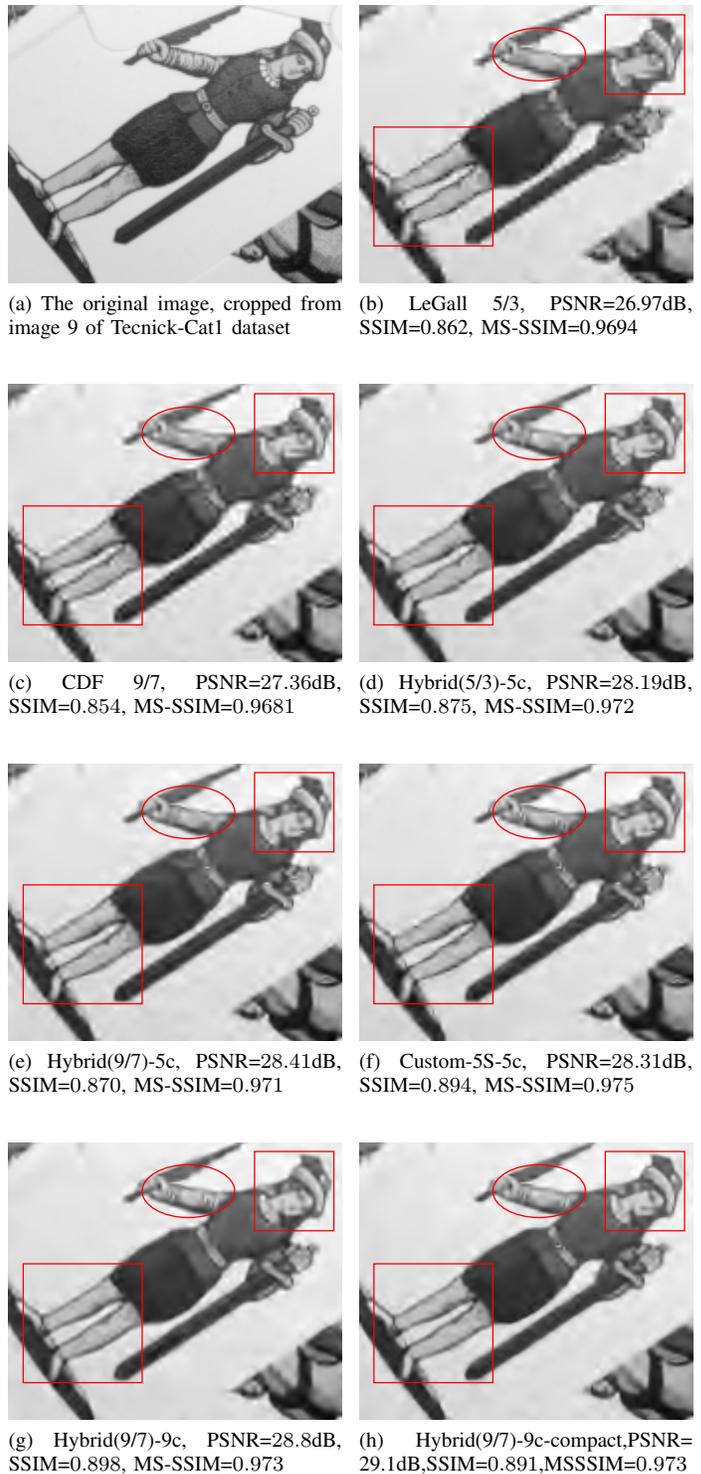

We can see that the hybrid lifting structure with two learned lifting steps still performs better than the other structures considered.
This conclusion is reinforced when we inspect the visual quality of fully reconstructed images, as seen in Fig.~\ref{fig:reconstructed_img}.
Moreover, Table~\ref{table:computational_complexity} demonstrates that the hybrid lifting structure also comes with modest region of support and computational complexity.

Of course, it is likely that a predict-update structure with even more learned lifting steps can outperform the hybrid lifting structure with the CDF 9/7 base wavelet transform.
However, this is certainly at the cost of high computational complexity, as well as a very large region of support; both damage the fundamental features of the JPEG 2000 codec, whose importance has been highlighted in Section~\ref{sec:intro}.
Moreover, other researchers have also explored lifting structures which employ longer sequences of learned lifting steps \cite{dardouri2020optimized, dardouri2021neural, dardouri2021dynamic}; in fact, results obtained from these works do not provide substantial improvements compared to JPEG 2000.

These observations suggest that to improve the conventional wavelet transform with neural networks, it may not be worthwhile to develop deep fully learned lifting structures.
Instead, it appears to be more beneficial to augment a larger base wavelet transform with two additional learned lifting steps $\mathcal{T}^A_{H2L}$ and $\mathcal{T}^A_{L2H}$.
This approach provides competitive coding performance across various evaluation metrics for different types of images, while exhibiting much lower computational complexity and more compact region of support.

\subsubsection{Merits of increasing the diversity of learned lifting steps}
\label{subsubsec:merits_diversity}
Now we move on to study the merit of increasing the diversity of learned lifting steps (i.e. the number of channels in each learned lifting operator) on coding performance; this does not have any impact on region of support, unlike increasing the depth of lifting structures.

\begin{figure}[tb]
	\centering
	\includegraphics[width=1.0\linewidth]{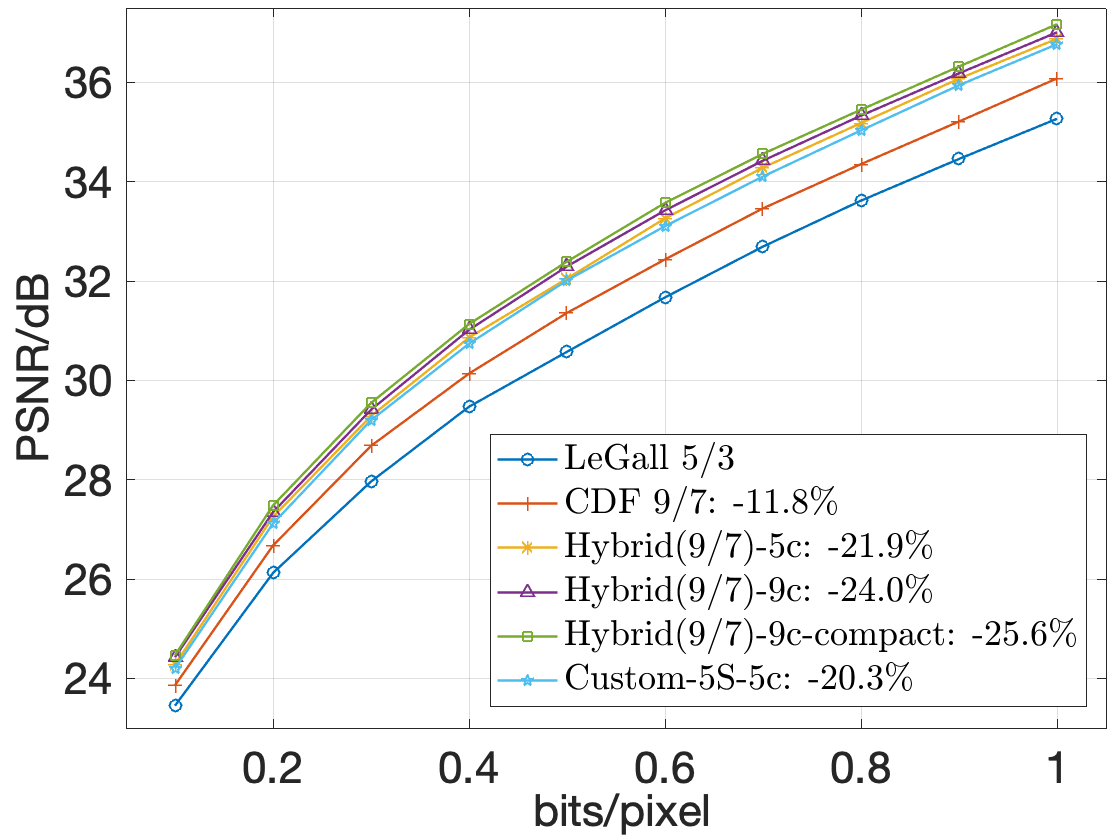}
	\caption{Comparisons of the average PSNR improvements over the LeGall 5/3 and CDF 9/7 wavelet transforms across \emph{Tecnick-Cat1} dataset, using different channels $N$ as well as different spatial support for each learned lifting operator. Bj{\o}ntegaard (BD) rate savings (in \%) are displayed next to legends; we use the LeGall 5/3 wavelet transform as the anchor here.}
	\label{fig:merits_diversity}
\end{figure}

Specifically, we start from the hybrid structure \emph{Hybrid(9/7)-5c}, whose superiority has been demonstrated in Section~\ref{subsubsec:merits_depth}.
Instead of using $N = 5$, we explore the benefit of using $N = 9$ channels; the resulting variation is referred to as \emph{Hybrid(9/7)-9c}.
The training procedure is identical to that of Section~\ref{subsubsec:merits_depth}.
The BD-rate savings (in \%) for average PNSR, SSIM and MS-SSIM over the range of bit-rates from $0.1$bpp to $1.0$bpp is provided in Table~\ref{table:merits_depth}.
Key examples of rate-distortion curves are shown in Fig.~\ref{fig:merits_diversity}.

We can see that increasing the diversity of learned lifting steps further improves coding efficiency of the corresponding lifting structure; the associated impact on computational complexity and region of support is shown in Table~\ref{table:computational_complexity}.
More interestingly, we observe that the hybrid lifting structure Hybrid(9/7)-9c exhibits the highest coding efficiency, especially for the PSNR metric, amongst other configurations with more learned lifting steps.
This configuration also produces visually appealing reconstructed images as illustrated in Fig.~\ref{fig:reconstructed_img}.

These observations align with the conclusion in Section~\ref{subsubsec:merits_depth} that it is more beneficial to employ learned $\mathcal{T}^A_{H2L}$ and $\mathcal{T}^A_{L2H}$ steps to improve the conventional base wavelet transform, rather than developing deep fully learned lifting structures.
Moreover, if we can afford additional computational complexity and are only interested in coding efficiency, rather than employing more learned lifting steps, we can choose to increase the diversity of $\mathcal{T}^A_{H2L}$ and $\mathcal{T}^A_{L2H}$ networks for higher compression performance.

\subsubsection{Study on spatial support of learned lifting steps}
\label{subsubsec:merits_width}

Compact region of support is one of the fundamental features of the conventional wavelet transform.
This feature, however, is damaged by augmenting or replacing the fixed lifting steps that correspond to the base wavelet transform with neural networks, which generally have substantially larger spatial supports. 
Therefore, it is important to study whether similar coding performance can be achieved using lifting networks with more compact region of support or not.

To study this, we start with the hybrid lifting structure \emph{Hybrid(9/7)-9c}, whose superiority over other configurations has been demonstrated in Section~\ref{subsubsec:merits_diversity}.
Specifically, we set all convolutional kernels in Hybrid(9/7)-9c to be $3$ x $3$ and removing the last residual block shown in Fig.~\ref{fig:proposed_H2L_L2H_networks}. 
The resulting lifting structure \emph{Hybrid(9/7)-9c-compact} then has significantly smaller expansion in the image domain with one level of decomposition, together with lower computational complexity than \emph{Hybrid(9/7)-9c}, as demonstrated in Table~\ref{table:computational_complexity}.

Table~\ref{table:merits_depth} and Fig.~\ref{fig:merits_diversity} show that competitive (even slightly better) coding performance can be achieved using more diverse $\mathcal{T}^A_{H2L}$ and $\mathcal{T}^A_{L2H}$ networks with more compact region of support.
This configuration also produces slightly enhanced full reconstructed images as shown in Fig.~\ref{fig:reconstructed_img}.
These observations reinforce the statement in Section~\ref{subsec:LS_more_steps} that it is important for a highly scalable compression system to have limited number of non-linearities; otherwise, quantization errors can expand in an uncontrollable way during synthesis.
Ultimately, \emph{Hybrid(9/7)-9c-compact} is our recommended approach to build learned wavelet-like transforms.
\begin{table}[htb!]
	\caption{Comparisons of computational complexity and region of support (expansion in the image domain with one level of decomposition)}
	\centering
	\resizebox{0.5\textwidth}{!}{
	\begin{tabular}{ccc}
		\toprule
		& Number of Parameters & Region of Support \\ \midrule
		\makecell{Hybrid(5/3)-5c}  & $35K$ & $78$ x $78$ \\
		\makecell{Hybrid(9/7)-5c}  & $35K$ & $82$ x $82$ \\
		\makecell{Hybrid(5/3)-9c}  & $63K$ & $78$ x $78$ \\
		\makecell{Hybrid(9/7)-9c}  & $63K$ & $82$ x $82$ \\
		\makecell{Hybrid(9/7)-9c-compact} & $35K$ & $54$ x $54$\\
		\makecell{Custom-4S-5c}  & $38K$ & $162$ x $162$ \\
		\makecell{Custom-4MS-5c}  & $55K$ & $202$ x $202$ \\
		\makecell{Custom-5S-5c}  & $73K$ & $234$ x $234$ \\
		\makecell{Custom-4S-9c}  & $69K$ & $162$ x $162$ \\
		\makecell{Custom-4MS-9c}  & $98K$ & $202$ x $202$ \\
		\makecell{Custom-5S-9c}  & $133K$ & $234$ x $234$ \\ \bottomrule
	\end{tabular}}
	\label{table:computational_complexity}
\end{table}

\section{Conclusion and Recommendations}
\label{sec:conclusion}
In this paper, we demonstrate the possibility to develop an end-to-end optimised image compression framework, which retains the resolution scalability, quality scalability and region of interest accessibility features from JPEG 2000, while producing significant gains in coding efficiency.

Restricting our attention to the transform structure, although it seems tempting to apply neural networks to all lifting steps of the wavelet transform to improve coding efficiency, this paper shows that it is more fruitful to augment a good fixed wavelet transform with two additional learned lifting steps.

As far as each individual learned lifting operator is concerned, the proposal-opacity network topology, which is comprised of linear proposals modulated by non-linear opacities, appears to have significant benefits over more traditional convolutional neural network topologies.
Ultimately, these proposal-opacity lifting networks can provide enhanced coding efficiency with increased number of channels and more compact region of support.

We remind the reader that the purpose of this study is \emph{only} to investigate the merits of different transform structures themselves.
For this very purpose, we deliberately choose not to incorporate machine learning into all aspects of the compression system, as they may obscure the contribution made solely by the transform.

\bibliographystyle{IEEEbib}
\bibliography{refs}

\end{document}